\documentclass[manuscript, letterpaper]{aastex6}

\pdfoutput=1

\newcommand{\githash}{a485057}\newcommand{\gitdate}{2016-09-30}

\newcommand{\numinner}{5}
\newcommand{\numkois}{844}
\newcommand{\numtargets}{39,036}
\newcommand{\numinjs}{819,752}
\newcommand{\numcands}{16}
\newcommand{\massocc}{\ensuremath{0.046\pm0.013}}
\newcommand{\massoccint}{\ensuremath{0.882\pm0.245}}
\newcommand{\masssemiocc}{\ensuremath{0.068\pm0.019}}
\newcommand{\masssemioccint}{\ensuremath{0.925\pm0.257}}
\newcommand{\intocc}{\ensuremath{2.00\pm0.72}}

\newcommand{\parama}{-0.20}
\newcommand{\paramb}{0.90}
\newcommand{\paramc}{-0.13}
\newcommand{\paramd}{0.95}
\newcommand{\parame}{0.70}
\newcommand{\paramf}{3.06}
\newcommand{\paramg}{-0.07}
\newcommand{\paramh}{-0.91}

\usepackage{microtype}

\usepackage{url}
\usepackage{amssymb,amsmath}
\usepackage{natbib}
\usepackage{multirow}
\hypersetup{colorlinks = false}
\makeatletter 
\long\def\frontmatter@title@above{
\vspace*{-\headsep}\vspace*{\headheight}
\noindent\footnotesize
{\noindent\footnotesize\textsc{\@journalinfo}}\par
{\noindent\scriptsize Preprint typeset using \LaTeX\ style AASTeX6\\
With modifications by David W. Hogg and Daniel Foreman-Mackey
}\par\vspace*{-\baselineskip}\vspace*{0.625in}
}%
\makeatother

\makeatletter
\let\origsection\section
\renewcommand\section{\@ifstar{\starsection}{\nostarsection}}
\newcommand\nostarsection[1]{\sectionprelude\origsection{#1}}
\newcommand\starsection[1]{\sectionprelude\origsection*{#1}}
\newcommand\sectionprelude{\vspace{1em}}
\let\origsubsection\subsection
\renewcommand\subsection{\@ifstar{\starsubsection}{\nostarsubsection}}
\newcommand\nostarsubsection[1]{\subsectionprelude\origsubsection{#1}}
\newcommand\starsubsection[1]{\subsectionprelude\origsubsection*{#1}}
\newcommand\subsectionprelude{\vspace{1em}}
\makeatother

\sloppypar

\newcommand{\project}[1]{\textsl{#1}}
\newcommand{\kepler}{\project{Kepler}}
\newcommand{\KT}{\project{K2}}
\newcommand{\tess}{\project{TESS}}
\newcommand{\plato}{\project{PLATO}}
\newcommand{\gaia}{\project{Gaia}}
\newcommand{\pdc}{\project{PDC}}
\newcommand{\bls}{\project{BLS}}
\newcommand{\emcee}{\project{emcee}}
\newcommand{\exosyspop}{\project{exosyspop}}

\newcommand{\foreign}[1]{\emph{#1}}
\newcommand{\etal}{\foreign{et\,al.}}
\newcommand{\etc}{\foreign{etc.}}

\newcommand{\dfmfigref}[1]{\ref{fig:#1}}
\newcommand{\dfmFig}[1]{Figure~\dfmfigref{#1}}
\newcommand{\dfmfig}[1]{\dfmFig{#1}}
\newcommand{\dfmfiglabel}[1]{\label{fig:#1}}

\renewcommand{\eqref}[1]{\ref{eq:#1}}
\newcommand{\Eq}[1]{Equation~(\eqref{#1})}
\newcommand{\eq}[1]{\Eq{#1}}
\newcommand{\eqalt}[1]{Equation~\eqref{#1}}
\newcommand{\eqlabel}[1]{\label{eq:#1}}

\newcommand{\sectionname}{Section}
\newcommand{\sectref}[1]{\ref{sect:#1}}
\newcommand{\Sect}[1]{\sectionname~\sectref{#1}}
\newcommand{\sect}[1]{\Sect{#1}}

\newcommand{\App}[1]{Appendix~\sectref{#1}}
\newcommand{\app}[1]{\App{#1}}
\newcommand{\sectlabel}[1]{\label{sect:#1}}

\newcommand{\T}{\ensuremath{\mathrm{T}}}
\newcommand{\dd}{\ensuremath{\,\mathrm{d}}}
\newcommand{\unit}[1]{{\ensuremath{\,\mathrm{#1}}}}
\newcommand{\bvec}[1]{{\ensuremath{\boldsymbol{#1}}}}
\newcommand{\appropto}{\mathrel{\vcenter{
  \offinterlineskip\halign{\hfil$##$\cr
    \propto\cr\noalign{\kern2pt}\sim\cr\noalign{\kern-2pt}}}}}

\newcommand{\response}[1]{{#1}}

\newcommand{\paper}{paper}

\newcommand{\meanpars}{{\ensuremath{\bvec{\theta}}}}
\newcommand{\kernpars}{{\ensuremath{\bvec{\alpha}}}}
\newcommand{\params}{{\ensuremath{\bvec{w}}}}

\newcommand{\modelname}[1]{{\textsf{#1}}}
\newcommand{\datareleaseurl}{\url{http://dx.doi.org/10.5281/zenodo.58273}}

\shorttitle{The population of long-period transiting exoplanets}
\shortauthors{Foreman-Mackey, Morton, Hogg, \etal}

\begin{document}

\title{%
\vspace{\baselineskip}
The population of long-period transiting exoplanets
\vspace{-2\baselineskip}  
}

\newcounter{affilcounter}
\altaffiltext{1}{\textsf{danfm@uw.edu}; Sagan Fellow}

\setcounter{affilcounter}{2}
\edef \uw {\arabic{affilcounter}}\stepcounter{affilcounter}
\altaffiltext{\uw}       {Astronomy Department, University of Washington,
                          Seattle, WA, 98195, USA}

\edef \princeton {\arabic{affilcounter}}\stepcounter{affilcounter}
\altaffiltext{\princeton}{Department of Astrophysics, Princeton University,
                          Princeton, NJ, 08544, USA}

\edef \scda {\arabic{affilcounter}}\stepcounter{affilcounter}
\altaffiltext{\scda}     {Simons Center for Data Analysis, 160 Fifth Avenue,
                          7th floor, New York, NY 10010, USA}

\edef \nyu       {\arabic{affilcounter}}\stepcounter{affilcounter}
\altaffiltext{\nyu}      {Center for Cosmology and Particle Physics,
                          New York University,
                          4 Washington Place, New York, NY, 10003, USA}

\edef \mpia      {\arabic{affilcounter}}\stepcounter{affilcounter}
\altaffiltext{\mpia}     {Max-Planck-Institut f\"ur Astronomie,
                          K\"onigstuhl 17, D-69117 Heidelberg, Germany}

\edef \cds       {\arabic{affilcounter}}\stepcounter{affilcounter}
\altaffiltext{\cds}      {Center for Data Science, New York University,
                          726 Broadway, 7th Floor, New York, NY, 10003, USA}

\edef \mpis      {\arabic{affilcounter}}\stepcounter{affilcounter}
\altaffiltext{\mpis}     {Max Planck Institute for Intelligent Systems
                          Spemannstrasse 38, 72076 T\"ubingen, Germany}

\author{%
    Daniel~Foreman-Mackey\altaffilmark{1,\uw},
    Timothy~D.~Morton\altaffilmark{\princeton},
    David~W.~Hogg\altaffilmark{\scda,\nyu,\mpia,\cds},
    Eric~Agol\altaffilmark{\uw}, and
    Bernhard~Sch\"olkopf\altaffilmark{\mpis}
    \vspace{\baselineskip}
}

\begin{abstract}

The \kepler\ Mission has discovered thousands of exoplanets and revolutionized
our understanding of their population.
This large, homogeneous catalog of discoveries has enabled rigorous studies of
the occurrence rate of exoplanets and planetary systems as a function of their
physical properties.
However, transit surveys like \kepler\ are most sensitive to planets with
orbital periods much shorter than the orbital periods of Jupiter and Saturn,
the most massive planets in our Solar System.
To address this deficiency, we perform a fully automated search for
long-period exoplanets with only one or two transits in the archival \kepler\
light curves.
When applied to the $\sim 40,000$ brightest Sun-like target stars, this search
produces \numcands\ long-period exoplanet candidates.
Of these candidates, 6 are novel discoveries and 5 are in systems with inner
short-period transiting planets.
Since our method involves no human intervention, we empirically characterize
the detection efficiency of our search.
Based on these results, we measure the average occurrence rate of exoplanets
smaller than Jupiter with orbital periods in the range 2--25~years to be
$2.0\pm0.7$ planets per Sun-like star.

\end{abstract}


\clearpage
\section{Introduction}

Data from the \kepler\ Mission \citep{Borucki:2011} have been used to discover
thousands of transiting exoplanets.
The systematic nature of these discoveries and careful quantification of
survey selection effects, search completeness, and catalog reliability has
enabled many diverse studies of the detailed frequency and distribution of
exoplanets \citep[for example,][]{Howard:2012, Petigura:2013,
Foreman-Mackey:2014, Dressing:2015, Burke:2015, Mulders:2015}.
So far, these results have been limited to relatively short orbital periods
because existing transit search methods impose the requirement of the detection of at least
three transits within the baseline of the data.
For \kepler, with a baseline of about four years, this sets an absolute upper
limit of about two years on the range of detectable periods.
In the Solar System, Jupiter~--~with a period of 12 years~--~dominates the
planetary dynamics and, since it would only exhibit at most one transit in the
\kepler\ data, an exo-Jupiter would be missed by most existing transit search
procedures.

Before the launch of the \kepler\ Mission, it was predicted that the nominal
mission would discover at least 10 exoplanets with only one or two observed
transits \citep{Yee:2008}, yet subsequent searches for these signals have
already been more fruitful than expected \citep{Wang:2015, Uehara:2016}.
However, the systematic study of the population of long-period exoplanets
found in the \kepler\ data to date has been hampered due to the
substantial technical challenge of implementing a search, as well as the
subtleties involved in interpreting the results. For example,
false alarms in the form of uncorrected systematics in the data and background
eclipsing binaries can make single-transit detections ambiguous.

Any single transit events discovered in the \kepler\ light curves are
interesting in their own right, but the development of a general and
systematic method for the discovery of planets with orbital periods longer
than the survey baseline is also crucial for the future of exoplanet research
with the transit method.
All future transit surveys have shorter observational baselines than the
\kepler\ Mission (\KT, \citealt{Howell:2014}; \tess, \citealt{Ricker:2015};
\plato, \citealt{Rauer:2014}) and given suitable techniques, single transit
events will be plentiful and easily discovered.
The methodological framework presented in the following pages is a candidate
for this task.

A study of the population of long-period transiting planets complements other
planet detection and characterization techniques, such as radial velocity
\citep[for example][]{Cumming:2008, Knutson:2014, Bryan:2016}, microlensing
\citep[for example][]{Gould:2010, Cassan:2012, Clanton:2014,
Shvartzvald:2016}, direct imaging \citep[for example][]{Bowler:2016}, and
transmission spectroscopy \citep[for example][]{Dalba:2015}.
The marriage of the radial velocity and transit techniques is particularly
powerful as exoplanets with both mass and radius measurements can be used to
study planetary compositions and the formation of planetary systems \citep[for
example][]{Weiss:2014, Rogers:2015, Wolfgang:2016}.
Unfortunately the existing catalog of exoplanets with measured densities is sparsely
populated at long orbital periods;  this makes discoveries with the transit method
at long orbital period compelling targets for follow-up observations.
Furthermore, even at long orbital periods, the \kepler\ light curves should be
sensitive to planets at the detection limits of the current state-of-the-art
radial velocity surveys.

There are two main technical barriers to a systematic search for single
transit events.
The first is that the transit probability for long-period planets is very low;
scaling as $\propto P^{-5/3}$ for orbital periods $P$ longer than the
baseline of contiguous observations.
Therefore, even if long-period planets are intrinsically common, they will
be underrepresented in a transiting sample.
The second challenge is that there are many signals in the observed light
curves caused by stochastic processes~--~both instrumental and
astrophysical~--~that can masquerade as transits.
Even when the most sophisticated methods for removing this variability are
used, false signals far outnumber the true transit events in any traditional
search.

At the heart of all periodic transit search procedures is a filtering step
based on ``box least squares'' \citep[\bls;][]{Kovacs:2002}.
This step produces a list of candidate transit times that is then vetted to
remove the substantial fraction of false signals using some combination of
automated heuristics and expert curation.
In practice, the fraction of false signals can be substantially reduced by
requiring that at least three self-consistent transits be observed
\citep{Petigura:2013, Burke:2014, Rowe:2015, Coughlin:2016}.

Relaxing the requirement of three transits requires a higher signal-to-noise
threshold per transit for validating candidate planets that display only one
to two transits.
Higher signal-to-noise allows matching the candidate transit to the expected
shape of a limb-darkened light curve, as well as ruling out various false
alarms.  This is analagous to microlensing surveys, for which a planet can only be
detected once, thus requiring high signal-to-noise for a reliable detection
\citep{Gould:2004}.

Recent work has yielded discoveries of long-period transiting planets with
only one or two high signal-to-noise transits identified in archival \kepler\
and \KT\ light curves by visual inspection \citep{Wang:2013, Kipping:2014a,
Wang:2015, Osborn:2016, Kipping:2016, Uehara:2016}.
These discoveries have already yielded some tantalizing insight into the
population of long-period transiting planets but, since these previous results
rely on human interaction, it is prohibitively expensive to reliably measure
the completeness of these catalogs.
As a result, the existing catalogs of long-period transiting planets cannot be
used to rigorously constrain the occurrence rate of long-period planets.

In this \paper, we develop a systematic method for reliably discovering the
transits of large, long-period companions in photometric time series
\emph{without human intervention}.
The method is similar in character to the recently published fully automated
method used to generate the official DR24 exoplanet candidate catalog from
\kepler\ \citep{Mullally:2016, Coughlin:2016}.
Since the search methodology is fully automated, we can robustly measure the
search completeness~--~using injection and recovery tests~--~and use these
products to place probabilistic constraints on the occurrence rate of
long-period planets.
We apply this method to a subset of the archival data from the \kepler\
Mission, present a catalog of exoplanet candidates, and estimate the
occurrence rate of long-period exoplanets.
We finish by discussing the potential effects of false positives, evaluating the
prospects for follow-up, and comparing our results to other studies
based on different planet discovery methods.

\section{A fully automated search method}\sectlabel{search}

To find long-period exoplanets in the \kepler\ light curves, we search for
individual, high signal-to-noise transit signals using a fully automated
procedure that can be broken into three main steps:
\begin{enumerate}
{\item an initial candidate search using a box-shaped matched filter,}
{\item light curve-level vetting (using automated model comparison) to remove
signals that don't have a convincing transit shape, and}
{\item pixel-level vetting to remove some astrophysical false positives.}
\end{enumerate}
The following sections describe each of these steps in more detail.

The model comparison step (step 2) is the key component of our method that
enables robust automation but it is also computationally expensive because we
must estimate the marginalized likelihoods of several different models
describing a transit and other processes that ``look'' like transits but are
actually caused by noise.
This step is conservative: unless a signal is a very convincing transit, it
won't pass the test.
In practice, this means that all but the highest signal-to-noise events will
be rejected at this step.
Therefore, in the inexpensive first step~--~the initial candidate search~--~we
can restrict the candidate list to high signal-to-noise events without a
substantial loss in detection efficiency.

\subsection{Step 1 -- Initial candidate events}\sectlabel{stepone}

It is not computationally feasible to run a full model comparison at every
conceivable transit time in the light curve so we must first find potentially
interesting events.
For our purposes, ``interesting'' means high signal-to-noise and previously
unknown.

To generate this list, we use a method much like the standard ``box least
squares'' \citep[\bls;][]{Kovacs:2002} procedure with a single (non-periodic)
box.
After masking any known transits, we filter the \pdc\ light curves
\citep{Stumpe:2012, Smith:2012} using a running windowed median with a
half-width of 2~days to remove stellar variability.
We then compute the signal-to-noise of the depth of a 0.6~day-long top hat on
a grid of times spanning the full baseline of observations.

In detail, at each proposal time $t_0$, we hypothesize a box-shaped transit
with duration $\tau$
\begin{eqnarray}
m(t) &=& \left\{\begin{array}{ll}
    \mu - \delta, & \mbox{\,if $|t - t_0| < \tau/2$} \\
    \mu, & \mbox{\,otherwise}
\end{array}\right. \quad.
\end{eqnarray}
Assuming that the uncertainties on the observed fluxes $f(t)$ are Gaussian
with known variance ${\sigma_f}^2$, the likelihood function for the mean flux
$\mu$ and transit depth $\delta$ can be analytically computed to be a
two-dimensional Gaussian with mean and covariance given by linear
least-squares.
This likelihood function provides a natural scalar objective: the
signal-to-noise of the measured depth computed as a function of time.
In principle this scalar is also a function of duration but we only use a
single transit duration because the following steps in this procedure are only
sensitive to transits with very high signal-to-noise, and in practice, the
final results are insensitive to the specific choice of duration.

To avoid edge effects, we apodize this detection scalar near any large gaps in
the time series using a logistic function with width equal to one transit
duration.
Finally, we estimate the background noise level in the detection scalar time
series using a robust running windowed variance estimate of the detection
scalar.  We accept peaks that are more than 25-times this background noise
level as candidates.

For the \kepler\ light curves, this procedure yields at least one candidate
event in about 1~percent of targets.  For these targets, we investigate
the three highest signal-to-noise events in the following step.

\subsection{Step 2 -- Light curve vetting}\sectlabel{light-curve-vetting}

In this step of the method, the goal is to discard any signals that are not
sufficiently ``transit-like'' in shape.
This step is similar to the method independently developed and recently
published by the \kepler\ team \citep{Mullally:2016}.
To quantify the quality of a candidate, we perform a model comparison between
a physical transit model and a set of other parameterized models for
systematics.
In order for a candidate to pass this vetting step, the transit model must be
``preferred'' to any other model as measured using the Bayesian Information
Criterion (BIC).
The BIC is not the optimal choice for this model comparison, but it is more
computationally tractable than the alternatives, such as computing thousands
of precise marginalized likelihoods or expected utilities for each model.
The BIC can be efficiently computed and it exhibits the desired
behavior~--~decreasing with increasing likelihood but flexible models are
penalized~--~and we find that it performs sufficiently well in practice.

For up to three candidate transit times per light curve, we select a
contiguous chunk of \pdc\ light curve approximately centered on the proposed
transit with no more than 500 cadences (about 10 days) and compute the BIC of
each model for this data set.
The BIC for a model $k$ in the set of $K$ models is given by
\begin{eqnarray}
\mathrm{BIC}_k &=& -2\,\ln \mathcal{L}^* + J\,\ln N
\end{eqnarray}
where the likelihood function $\mathcal{L}$ is evaluated at its maximum, $J$
is the number of free parameters in the model, and $N$ is the number of
data points in the data set.

For each model, we describe the data using a Gaussian Process
\citep[GP;][]{Rasmussen:2006} with a Mat\'ern-3/2 covariance and mean given by
the chosen model $m_k(t;\,\meanpars)$ parameterized by the parameter vector
\meanpars.

We consider the following mean models (this list provides a qualitative
justification for each model):
\begin{itemize}
{\item \modelname{transit} -- a limb-darkened, exposure-time integrated
transit light curve,}
{\item \modelname{variability} -- a pure GP model to capture
stellar variability,}
{\item \modelname{outlier} -- a single outlier to account for a bad data
point,}
{\item \modelname{step} -- a step function to describe sudden pixel
sensitivity dropouts \citep[SPSDs; for example][]{Christiansen:2013}, and}
{\item \modelname{box} -- a box to catch signals that are well fit by the
search scalar but insufficiently transit-like to be convincing.}
\end{itemize}
The functional forms of these models are given in \app{model-details} and the
details of the technical methodology of GP fitting are described in
\app{gp-regression}.

\dfmfig{model-comp} shows representative events that fall into different
classes and the corresponding maximum likelihood model.
For each candidate event, the BIC of each of these models is computed and the
event is only passed as a candidate if the \modelname{transit} model is
preferred to all the other models.
The \modelname{box} model is the most restrictive comparison, vetoing about
half of the candidate events in the \kepler\ light curves, followed by the
\modelname{variability} model.
To further restrict to non-grazing transits, we also reject events where the
maximum likelihood impact parameter is greater than $1 - R_\mathrm{P} /
R_\star$.

Since the search procedure described here was tuned to discover transit
signals, we do not consider the distribution or potential astrophysical nature
of any models besides the \modelname{transit} model.
In the future, it would be interesting to relax this goal and investigate the
other model classes; in particular, the \modelname{box} model is
sensitive to astrophysical phenomena, notably occultations of white dwarfs.
In a cursory investigation it is clear that the majority of signals labeled
\modelname{box} in our analysis are noise; however, a subset are likely to be
astrophysical in nature.

The reliability of this method of automated vetting is limited by the specific
models selected in this step.
We find that these are sufficient for the targets discussed below but
different target lists or data sets might require additional models to be
included for robust selection.

\begin{figure*}[p]~\\
\begin{center}
\includegraphics[width=\textwidth]{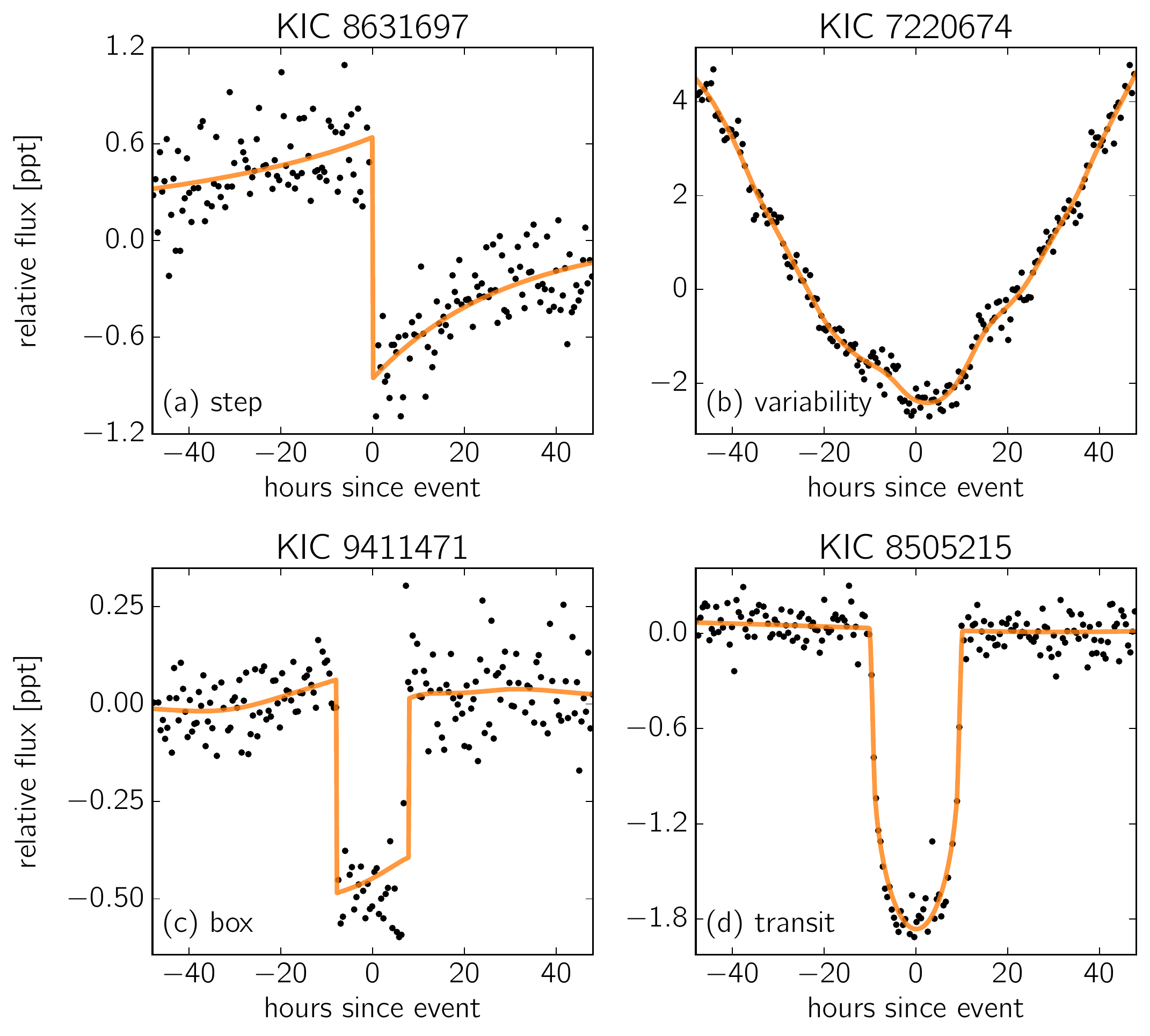}
\end{center}
\caption{%
Representative examples of candidate events flagged by the initial search.
Each example falls into a different model category and the figure shows the
data as black points and the best fit mean model prediction.
The examples represent the following model categories:
\emph{(a)} step \emph{(b)} variability, \emph{(c)} box, and \emph{(d)}
transit.
\dfmfiglabel{model-comp}}
\end{figure*}

\subsection{Step 3 -- Pixel-level vetting}

To minimize contamination from background eclipsing binary systems, we require
candidate events to pass a centroid shift test similar to the one used in the
official \kepler\ transit search pipeline \citep{Bryson:2013}.
To measure the centroid shift, we model the flux-weighted centroid traces
independently in each coordinate as a multiple of the best-fit transit model
and a GP noise model.
By properly normalizing the transit model, we  measure the in-transit centroid
shift $\Delta_\mathrm{centroid}$ in pixels.
We reject any candidate event where the estimated transit location is more
than half a pixel from the out-of-transit centroid
\begin{eqnarray}
\Delta_\mathrm{centroid}\,\left(\frac{1}{\delta} - 1\right) &>& 0.5
\end{eqnarray}
where $\delta$ is the observed transit depth \citep{Bryson:2013}.

\section{Results: a catalog of long-period transiting exoplanet candidates}

To limit the scope of this paper while still demonstrating the applicability
of our method, we search the \kepler\ archival light curves of the brightest
and quietest Sun-like stars for long-period transiting exoplanets.
In this section, we describe the target selection process and the parameter
estimation procedure.

\subsection{Target selection}\sectlabel{target-selection}

We select the $\sim40,000$ brightest and quietest G and K dwarfs from the
\kepler\ catalog using the most recent catalog of stellar parameters%
\footnote{Parameters from the \textsf{q1\_q17\_dr24\_stellar} table from the
NASA Exoplanet Archive \citep[][with updates]{Huber:2014}.} and the cuts used
by \citet{Burke:2015}:
\begin{itemize}
{\item $4200\unit{K} \le T_\mathrm{eff} \le 6100\unit{K}$,}
{\item $R_\star \le 1.15\,R_\odot$,}
{\item $\mathrm{data\,span} \ge 2\,\mathrm{years}$,}
{\item $\mathrm{duty\,cycle} \ge 0.6$,}
{\item $K_p \le 15\unit{mag}$, and}
{\item $\mathrm{CDPP}_{7.5\unit{hrs}} \le 1000\unit{ppm}$.}
\end{itemize}
We continue by excluding the light curves of known eclipsing
binaries\footnote{\url{http://keplerebs.villanova.edu/}} \citep{Kirk:2016},
other known false positives \citep{Coughlin:2016}, a planet with known transit
timing variations (Kepler-9), and four especially noisy stars (KIC~4482348,
KIC~4450472, KIC~5438845, and KIC~10068041).
The final catalog contains \numtargets\ targets and the parameter distribution
is shown in \dfmfig{targets}.

Since these data have already been searched for short-period planets, we
assume that all high signal-to-noise candidates with three or more transits
have been previously found \citep{Coughlin:2016}.
To remove these candidates from consideration, we mask the cadences within two
transit durations of the time when a short-period planet candidate is known to
transit\footnote{We specifically use the \textsf{q1\_q17\_dr24\_koi} from the
NASA Exoplanet Archive \url{http://exoplanetarchive.ipac.caltech.edu/}.}.

\begin{figure}~\\
\begin{center}
\includegraphics{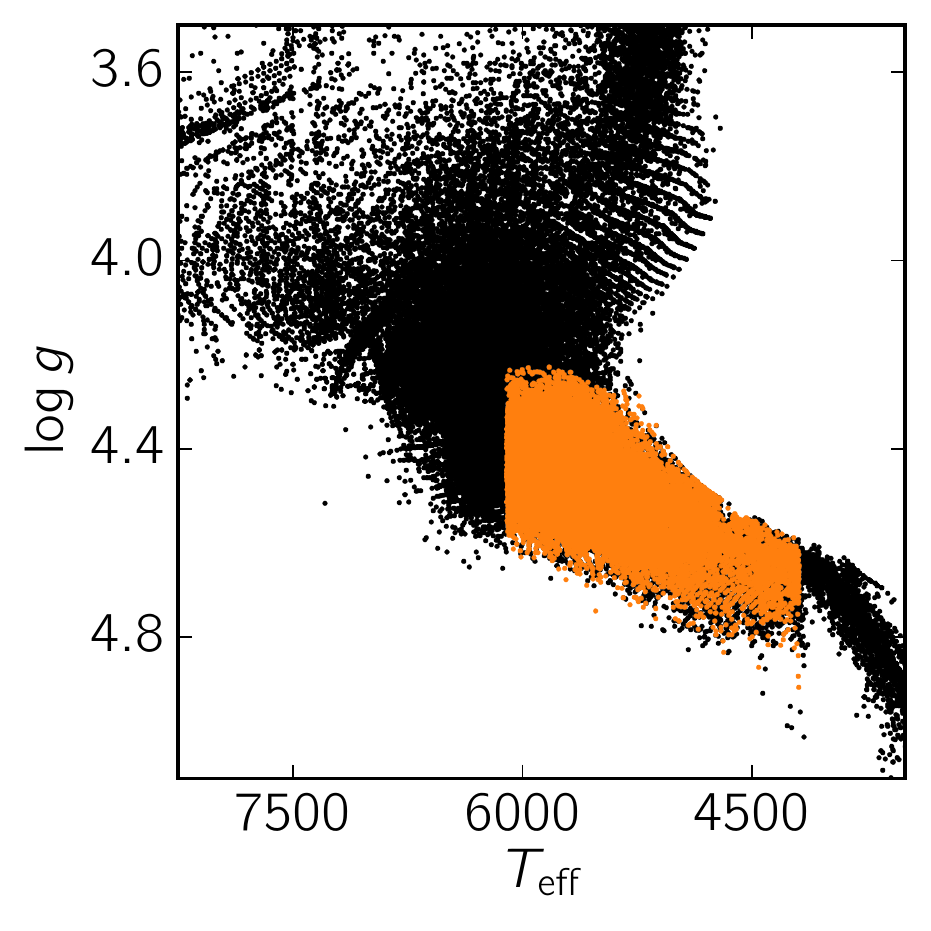}
\end{center}
\caption{%
The distribution of stellar parameters for \kepler\ targets selected for this
search (orange) compared to the distribution of the full \kepler\ target
catalog (black).
\dfmfiglabel{targets}}~\\
\end{figure}

\subsection{Parameter estimation}

For each transit candidate, we constrain the physical parameters of the system
by fitting a section of light curve around each transit using an exposure-time
integrated  Keplerian orbit with a quadratic limb darkening law for the
central body \cite{Foreman-Mackey:2016a}.
It has previously been established that the orbital period of a transiting
planet with only one transit can still be constrained given a measurement of
the stellar density and an assumption about the orbital eccentricity \cite[for
example][]{Wang:2015, Osborn:2016}.
Qualitatively this works because the transit of a bound body cannot have an
arbitrary period for a given duration.
This is the same argument used to justify the ``photoeccentric effect''
\citep{Dawson:2012} and the method of ``asterodensity profiling''
\citep{Kipping:2014b}.
In particular, this suggests that the periods of single transits in systems
with multiple inner planets will be especially well constrained
\citep{Kipping:2012}.
In this \paper, we do not take advantage of the extra constraints provided by
the inner planets, instead treating each long-period transiting system in
isolation, but this would be a good follow-up project.

In the following paragraphs, we describe the components of the probabilistic
model used to infer the planet candidates' properties.
To perform parameter estimation under this model, we use the Markov Chain
Monte Carlo (MCMC) package \emcee\footnote{\url{http://dfm.io/emcee}}
\citep{Foreman-Mackey:2013} with an ensemble of 40 walkers.
We run each chain until at least 750 independent samples~--~in most cases, we
actually produce thousands of independent samples~--~are obtained\footnote{The
integrated autocorrelation time is estimated using a robust iterative method
as suggested by Alan Sokal:
\url{http://www.stat.unc.edu/faculty/cji/Sokal.pdf}.} and discard the first
third of the chain as burn-in.
The posterior constraints on a few physical parameters for the single transit
candidate in the light curve of KIC~8505215 are shown in \dfmfig{corner} and
all the chains are made available online\footnote{\datareleaseurl}.

\begin{figure}~\\
\begin{center}
\includegraphics[width=\textwidth]{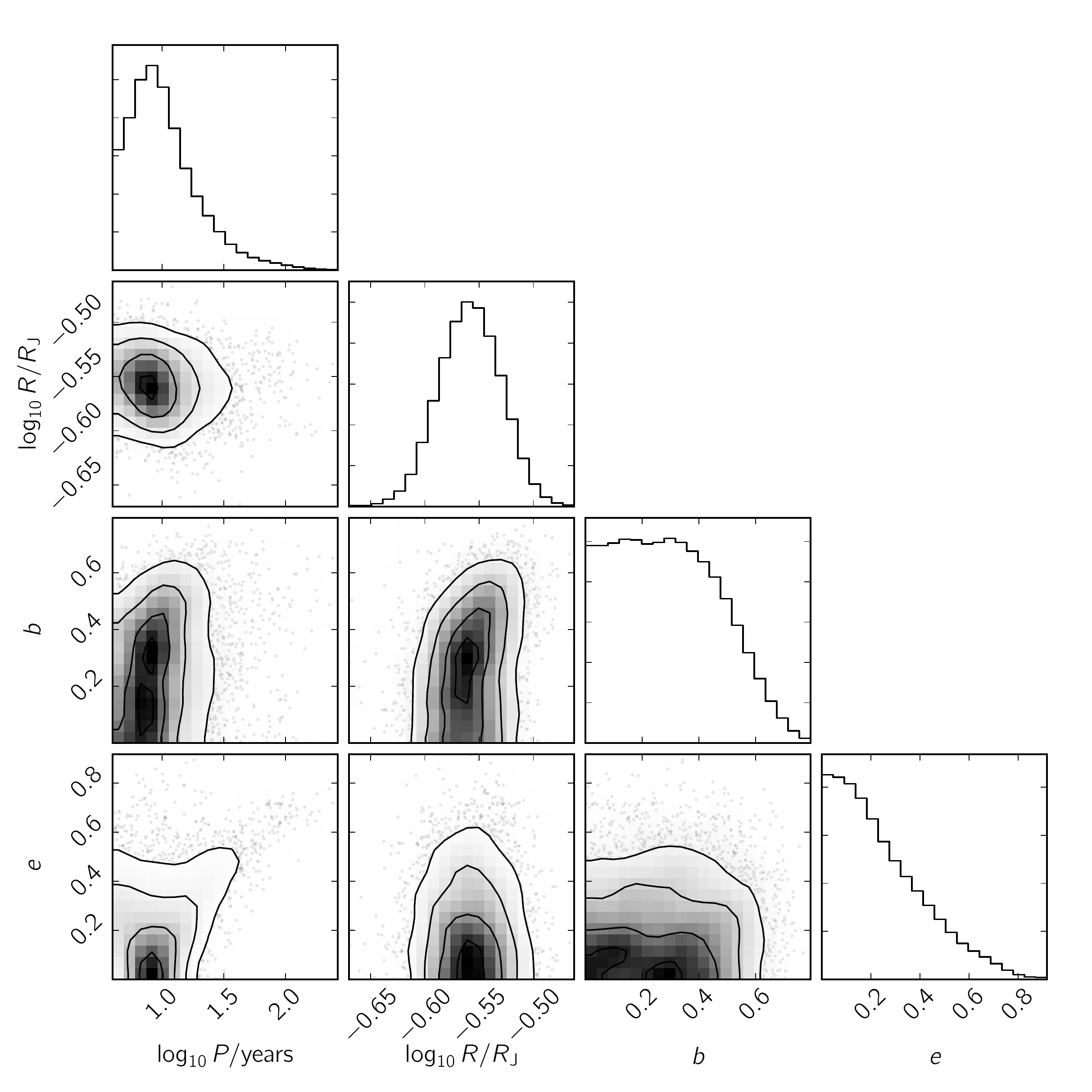}
\end{center}
\caption{%
The posterior constraints on the physical parameters for the single transit
candidate found in the light curve of KIC~8505215.
The contour plots show estimates of the two-dimensional marginalized
probability densities and the histograms show the marginalized density for
each parameter.
This figure was generated using \project{corner.py}
\citep{Foreman-Mackey:2016}.
\dfmfiglabel{corner}}~\\
\end{figure}

\paragraph{Priors}

For each candidate in our sample, we take the constraints on the stellar
parameters from the \kepler\ DR24 stellar properties catalog and assume an
empirical beta function prior on the eccentricities based on the observed
eccentricity distribution of long-period planets discovered using radial
velocities \citep{Kipping:2013}.
Table~\ref{tab:parameters} lists all the fit parameters and their prior
distributions.
Besides these listed priors, we add the extra constraint that no other
transits can occur in the baseline of the \kepler\ observations.
This constraint is overly conservative because there is some probability that
a second transit could occur in a data gap but we find that, in practice, most
of the posterior mass is at longer periods and the period inferences are not
significantly affected.

\paragraph{Likelihood function}

As above, we model the light curve as a Gaussian Process (GP) with a physical
transit model as the mean, and a covariance matrix described by a Mat\'ern-3/2
kernel function.
The full likelihood function and some details of GP regression are given in
\app{gp-regression}.
For computational efficiency, we first perform a joint optimization of the
physical parameters and GP hyperparameters to find the maximum \foreign{a
posteriori} model then keep the hyperparameters fixed and run MCMC sampling
for the 11 physical parameters alone.

\begin{floattable}
\begin{deluxetable}{lccc}
\tabletypesize{\footnotesize}
\caption{The inferred parameters and priors used in the inference
\label{tab:parameters}}

\tablehead{%
    \colhead{name} & \colhead{symbol} & \colhead{units} &
    \colhead{prior}
}

\startdata
mean flux & $\log f_\star$ & \nodata &
    $\log f_\star \sim \mathcal{U}(-1,\,1)$ \\
stellar mass\tablenotemark{a} & $M_\star$ & $M_\odot$ &
    $M_\star \sim \mathcal{N}(M_{\star,\mathrm{cat}},\,
        \sigma_{M,\star,\mathrm{cat}})$ \\
stellar radius\tablenotemark{a} & $R_\star$ & $R_\odot$ &
    $R_\star \sim \mathcal{N}(R_{\star,\mathrm{cat}},\,
        \sigma_{R,\star,\mathrm{cat}})$ \\
\multirow{2}{*}{limb darkening} & $q_1$ & \nodata &
    $q_1 \sim \mathcal{U}(0,\,1)$ \\
 & $q_2$ & \nodata & $q_2 \sim \mathcal{U}(0,\,1)$ \\
\hline
planet radius & $\log R_\mathrm{P}$ & $R_\odot$ &
    $\log R_\mathrm{P} \sim \mathcal{U}(-10,\,2)$ \\
reference time & $t_0$ & days &
    $t_0 \sim \mathcal{U}(t_\mathrm{cand}-0.5,\,t_\mathrm{cand}+0.5)$%
    \tablenotemark{b} \\
\multirow{2}{*}{\begin{minipage}{1in}semi-major axis \ \& inclination\end{minipage}}
    & $\sqrt{a}\sin i$ & ${R_\odot}^{1/2}$ &
    $\sqrt{a}\sin i \sim \mathcal{U}(-10^3,\,10^3) / \sqrt{a}$ \\
    & $\sqrt{a}\cos i$ & ${R_\odot}^{1/2}$ &
    $\sqrt{a}\cos i \sim \mathcal{U}(0,\,10^3) / \sqrt{a}$ \\
\multirow{2}{*}{eccentricity}
    & $\sqrt{e}\sin \omega$ & \nodata & $e \sim \beta(1.12,\,3.09)$%
    \tablenotemark{c} \\
    & $\sqrt{e}\cos \omega$ & \nodata & $\omega \sim \mathcal{U}(-\pi,\,\pi)$\\
\enddata

\tablenotetext{a}{Stellar parameters and uncertainties taken from the \kepler\
catalog \citep{Huber:2014}}
\tablenotetext{b}{The reference time is constrained to be within half a day of
the candidate transit time}
\tablenotetext{c}{\citet{Kipping:2013a}}
\tablecomments{There is one further constraint that complicates these priors:
the period of the orbit must be longer than some minimum period
$P_\mathrm{min}$ set by the transit time and the full baseline of \kepler\
observations.}
\end{deluxetable}
\end{floattable}

\section{Catalog of transit candidates}\sectlabel{catalog}

Applying the search procedure described in \sect{search} to the \kepler\ light
curves of the \numtargets\ targets selected in \sect{target-selection}, we
find \numcands\ convincing transit candidates.
Visual inspection of each candidate confirms the reliability of the
classification and no candidates are manually removed from the catalog.
Of these, three candidates have two transits in the \kepler\ baseline and the
remainder have only one observable transit.
The candidates and their inferred physical parameters are listed in
Table~\ref{tab:catalog} and the light curves are plotted in
\dfmfig{light-curves}.
The inferred radius and orbital periods of the candidates are compared to the
short-period \kepler\ sample and the Solar System in \dfmfig{full-sample}.

Two of the shortest period candidates~--~both with two observed
transits~--~have previously been studied in detail \citep[KIC~8800954 and
KIC~3239945;][]{Kipping:2014a, Kipping:2016}.
Table~\ref{tab:catalog} indicates the candidates that were also discovered by
earlier searches for long-period transiting systems using visual inspection
\citep{Wang:2015, Uehara:2016}.
The consistency between our results and the earlier catalogs is reassuring.
In the light curves of targets with previously known short-period planets, our
automated search did not find any candidates that weren't previously detected
by visual inspection \citep{Uehara:2016} and one candidate (KIC~3230491)
reported by the human analysis was discarded as grazing by our search.
The Planet Hunters citizen science project \citep{Fischer:2012} reported five
long-period candidates with one or two observed transits in our target list
\citep{Wang:2015}.
Of these, we also find two (KIC~8410697 and KIC~10842718) although we find a
second transit in the KIC~8410697 system that was previously missed.
We do not recover the three other candidates reported by \citet{Wang:2015}:
KIC~5536555, KIC~9662267, and KIC~12454613.
The transits of these candidates are all low signal-to-noise and they do not
pass our initial signal-to-noise threshold.
Six of the candidates in Table~\ref{tab:catalog} have not been previously
published.

Of the \numcands\ candidates, \numinner\ have known inner planets with three
or more observable transits \citep{Coughlin:2016}.
Given the fact that only \numkois\ of the \numtargets\ targets had previously
known planets, this means that systems with short-period transiting planets
are nearly a factor of 20 more likely to host long-period planets accessible
by our method than systems with no known inner transiting planets.
This difference cannot be accounted for by differences in completeness between
targets with known planets and without because the average detection
efficiency for both populations is consistent within sampling uncertainty.
Qualitatively, this suggests that these long-period planets occur with a
higher frequency in multi-planet systems or are preferentially aligned with
the plane of any inner planets but a more detailed analysis would be needed to
make a quantitative statement \citep[see, for example,][]{Tremaine:2012,
Fang:2012, Ballard:2016, Moriarty:2015}.

The candidate in the light curve of KIC~4754460 is an individual transit
candidate but another deeper eclipse can be found at a \kepler\ Barycentric
Julian Date (KBJD) of 1587.13; right at the beginning of Quarter 17.
This eclipse was missed by the automated search because only the second half
of the eclipse is observed.
The most likely explanation of this system is that the listed candidate is the
secondary eclipse of a binary system but we will keep the candidate in the
list and treat this effect statistically in \sect{false-positives}.

Five candidate transit events in the light curves of four targets were
rejected because of a significant centroid shift or a large impact parameter.
These events are probably astrophysical eclipses from binary star systems that
were not found by previous studies of long-period eclipsing binary systems.
We do not consider these events further in the following analysis but
Table~\ref{tab:rejects} lists these events and their properties for posterity.

\begin{floattable}
\begin{deluxetable}{cccccccccccl}
\tabletypesize{\scriptsize}
\caption{The inferred parameters for the long-period transiting exoplanet
candidates \label{tab:catalog}}
\tablehead{
    \colhead{kic id} &
    \colhead{$T_\mathrm{eff}$} &
    \colhead{$R_\star$} &
    \colhead{Kp} &
    \colhead{period} & \colhead{$t_0$} &
    \colhead{radius} &
    \colhead{duration} &
    \colhead{impact} &
    \colhead{$T_\mathrm{eq}$\tablenotemark{*}} &
    \colhead{$\mathrm{Pr}_\mathrm{planet}$} &
    \colhead{KOI/Kepler\tablenotemark{$\dagger$}}
    \\
    & \colhead{K} & \colhead{$R_\odot$} && \colhead{years} & \colhead{KBJD} &
    \colhead{$R_\mathrm{J}$} & \colhead{hours} && \colhead{K} &&
}
\rotate
\startdata
3218908$^{\mathrm{b}}$ & $5513_{-139}^{+172}$ & $0.75_{-0.05}^{+0.26}$ & $14.6$ & $7.0_{-3.4}^{+9.5}$ & $766.6722_{-0.0114}^{+0.0096}$ & $0.514_{-0.093}^{+0.092}$ & $21.45_{-0.61}^{+0.72}$ & $0.24_{-0.17}^{+0.21}$ & $129_{-39}^{+41}$ & $0.96$ & 1108 / 770\\
3239945$^{\mathrm{c}}$ & $4786_{-88}^{+106}$ & $0.71_{-0.03}^{+0.05}$ & $14.0$ & $2.9328721_{-0.0000026}^{+0.0000026}$ & $420.28714_{-0.00068}^{+0.00069}$ & $0.876_{-0.039}^{+0.039}$ & $16.202_{-0.071}^{+0.077}$ & $0.207_{-0.110}^{+0.071}$ & $142.8_{-4.3}^{+4.3}$ & $0.73$ & 490 / 167\\
4754460$^{\mathrm{}}$ & $5766_{-158}^{+172}$ & $1.13_{-0.22}^{+0.38}$ & $14.9$ & $5.9_{-3.0}^{+11.8}$ & $826.8369_{-0.0046}^{+0.0046}$ & $0.67_{-0.15}^{+0.16}$ & $15.92_{-0.54}^{+0.55}$ & $0.893_{-0.037}^{+0.018}$ & $171_{-64}^{+60}$ & $0.95$ & \\
6551440$^{\mathrm{}}$ & $6050_{-182}^{+155}$ & $1.10_{-0.15}^{+0.43}$ & $13.6$ & $4.0_{-1.2}^{+4.2}$ & $1039.0589_{-0.0037}^{+0.0037}$ & $0.282_{-0.083}^{+0.093}$ & $10.85_{-0.30}^{+0.37}$ & $0.60_{-0.37}^{+0.20}$ & $170_{-45}^{+38}$ & $0.97$ & \\
8410697$^{\mathrm{c,a}}$ & $5918_{-152}^{+157}$ & $1.00_{-0.16}^{+0.35}$ & $13.4$ & $2.8688097_{-0.0000054}^{+0.0000053}$ & $542.1231_{-0.0013}^{+0.0013}$ & $0.698_{-0.078}^{+0.107}$ & $19.77_{-0.10}^{+0.12}$ & $0.15_{-0.11}^{+0.12}$ & $206_{-13}^{+15}$ & $0.95$ & \\
8426957$^{\mathrm{}}$ & $5992_{-183}^{+153}$ & $1.08_{-0.13}^{+0.46}$ & $13.6$ & $54_{-36}^{+88}$ & $784.677_{-0.013}^{+0.013}$ & $1.04_{-0.25}^{+0.30}$ & $39.4_{-1.4}^{+1.6}$ & $0.889_{-0.059}^{+0.037}$ & $85_{-28}^{+46}$ & $0.80$ & \\
8505215$^{\mathrm{b}}$ & $5087_{-98}^{+102}$ & $0.71_{-0.03}^{+0.06}$ & $13.0$ & $9.1_{-3.4}^{+9.5}$ & $140.0492_{-0.0018}^{+0.0017}$ & $0.277_{-0.017}^{+0.017}$ & $20.06_{-0.16}^{+0.18}$ & $0.28_{-0.19}^{+0.20}$ & $103_{-23}^{+19}$ & $0.96$ & 99 / none\\
8738735$^{\mathrm{b}}$ & $6000_{-129}^{+101}$ & $1.04_{-0.06}^{+0.21}$ & $13.9$ & $9.9_{-5.0}^{+14.9}$ & $697.8538_{-0.0049}^{+0.0059}$ & $0.355_{-0.044}^{+0.045}$ & $27.44_{-0.37}^{+0.62}$ & $0.28_{-0.20}^{+0.30}$ & $137_{-39}^{+43}$ & $0.97$ & 693 / 214\\
8800954$^{\mathrm{c}}$ & $5286_{-101}^{+110}$ & $0.76_{-0.04}^{+0.06}$ & $13.4$ & $1.9279957_{-0.0000091}^{+0.0000092}$ & $492.7652_{-0.0024}^{+0.0024}$ & $0.386_{-0.025}^{+0.025}$ & $15.76_{-0.13}^{+0.14}$ & $0.18_{-0.13}^{+0.17}$ & $189.4_{-7.4}^{+7.2}$ & $0.95$ & 1274 / 421\\
9306307$^{\mathrm{}}$ & $5762_{-160}^{+209}$ & $0.92_{-0.15}^{+0.44}$ & $14.0$ & $4.3_{-1.1}^{+3.3}$ & $1191.35648_{-0.00018}^{+0.00018}$ & $1.22_{-0.36}^{+0.49}$ & $8.499_{-0.042}^{+0.042}$ & $0.6399_{-0.0072}^{+0.0068}$ & $126_{-32}^{+33}$ & $8.7 \times 10^{-6}$ & \\
10187159$^{\mathrm{b}}$ & $5185_{-143}^{+179}$ & $0.91_{-0.11}^{+0.80}$ & $14.4$ & $4.9_{-1.8}^{+7.6}$ & $604.1102_{-0.0031}^{+0.0023}$ & $0.43_{-0.13}^{+0.21}$ & $11.81_{-0.20}^{+0.27}$ & $0.23_{-0.16}^{+0.21}$ & $119_{-43}^{+34}$ & $0.91$ & 1870 / 989\\
10287723$^{\mathrm{b}}$ & $4500_{-119}^{+153}$ & $0.73_{-0.05}^{+0.04}$ & $13.4$ & $4.9_{-1.3}^{+4.3}$ & $393.5976_{-0.0029}^{+0.0031}$ & $0.266_{-0.024}^{+0.027}$ & $9.49_{-0.30}^{+0.33}$ & $0.68_{-0.36}^{+0.13}$ & $114_{-22}^{+13}$ & $0.95$ & 1174 / none\\
10321319$^{\mathrm{}}$ & $5749_{-133}^{+154}$ & $0.94_{-0.13}^{+0.38}$ & $11.9$ & $5.5_{-2.1}^{+8.1}$ & $554.3562_{-0.0063}^{+0.0064}$ & $0.163_{-0.037}^{+0.046}$ & $16.84_{-0.38}^{+0.42}$ & $0.47_{-0.31}^{+0.26}$ & $153_{-47}^{+37}$ & $1.00$ & \\
10602068$^{\mathrm{}}$ & $5628_{-157}^{+165}$ & $0.91_{-0.08}^{+0.36}$ & $14.9$ & $3.16_{-0.83}^{+2.65}$ & $830.80892_{-0.00015}^{+0.00015}$ & $2.00_{-0.35}^{+0.66}$ & $12.804_{-0.092}^{+0.068}$ & $0.6027_{-0.0078}^{+0.0072}$ & $159_{-36}^{+28}$ & $3.9 \times 10^{-9}$ & \\
10842718$^{\mathrm{a}}$ & $5754_{-156}^{+159}$ & $1.04_{-0.14}^{+0.38}$ & $14.6$ & $12.7_{-6.6}^{+20.2}$ & $226.2344_{-0.0047}^{+0.0047}$ & $0.74_{-0.16}^{+0.16}$ & $35.92_{-0.38}^{+0.51}$ & $0.26_{-0.17}^{+0.19}$ & $128_{-43}^{+47}$ & $0.90$ & \\
11709124$^{\mathrm{b}}$ & $5688_{-101}^{+108}$ & $0.97_{-0.09}^{+0.19}$ & $14.5$ & $4.3_{-1.3}^{+4.7}$ & $657.2674_{-0.0016}^{+0.0018}$ & $0.83_{-0.11}^{+0.12}$ & $17.75_{-0.44}^{+0.54}$ & $0.51_{-0.22}^{+0.12}$ & $166_{-39}^{+28}$ & $0.94$ & 435 / 154\\
\enddata

\tablenotetext{*}{The equilibrium temperature is computed assuming zero
albedo.}
\tablenotetext{\dagger}{The KOI number and, if applicable, the \kepler\ number
for the target.}
\tablenotetext{a}{Included in the \citet{Wang:2015} catalog.}
\tablenotetext{b}{Included in the \citet{Uehara:2016} catalog.}
\tablenotetext{c}{Candidate has two observed transits.}
\tablecomments{The values and uncertainties indicate the 16-th, 50-th, and
84-th percentiles of the posterior samples for each parameter.}
\end{deluxetable}
\end{floattable}

\begin{figure*}[p]~\\
\begin{center}
\includegraphics[width=0.24\textwidth]{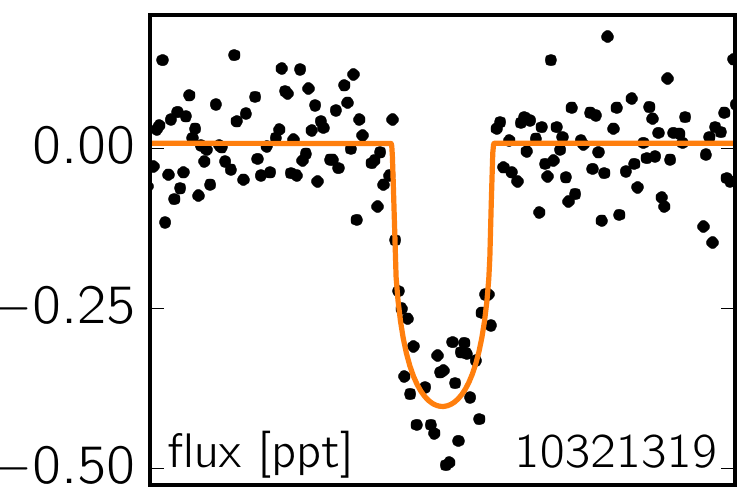}
\includegraphics[width=0.24\textwidth]{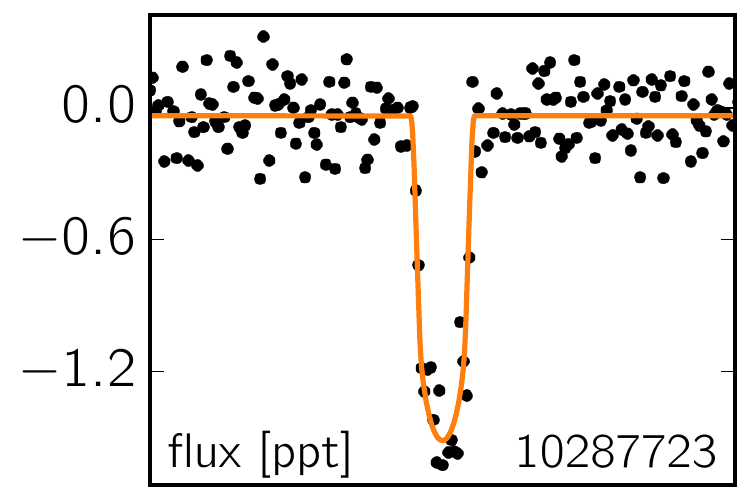}
\includegraphics[width=0.24\textwidth]{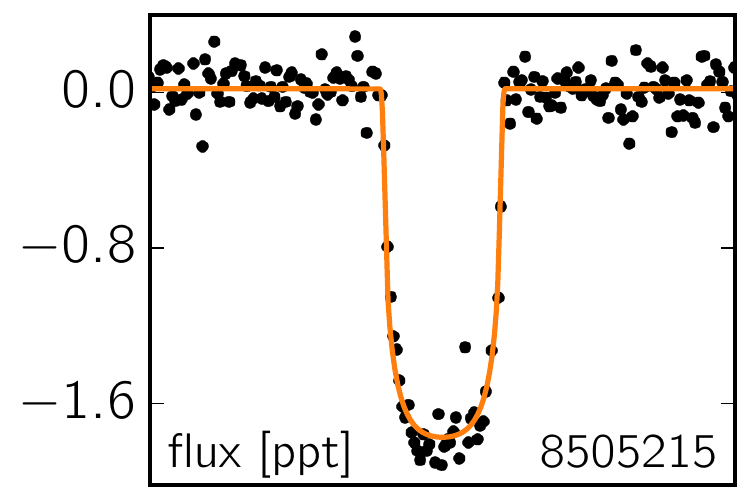}
\includegraphics[width=0.24\textwidth]{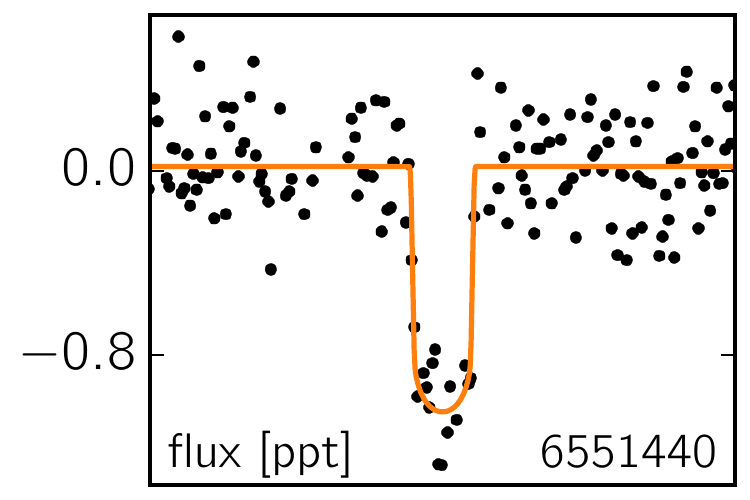}
\includegraphics[width=0.24\textwidth]{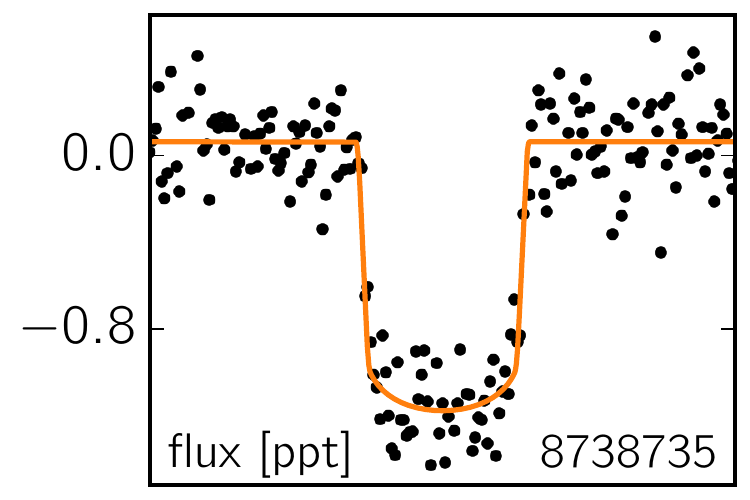}
\includegraphics[width=0.24\textwidth]{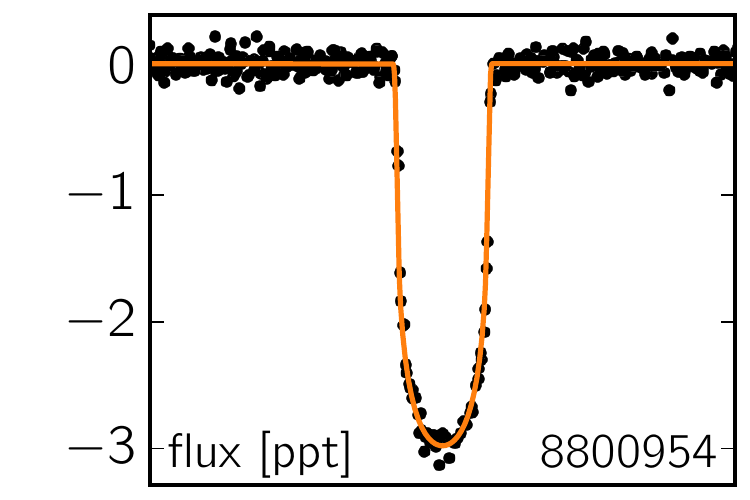}
\includegraphics[width=0.24\textwidth]{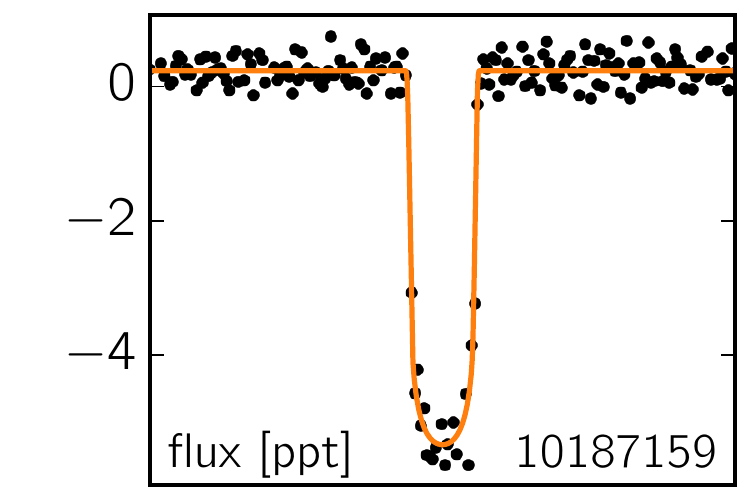}
\includegraphics[width=0.24\textwidth]{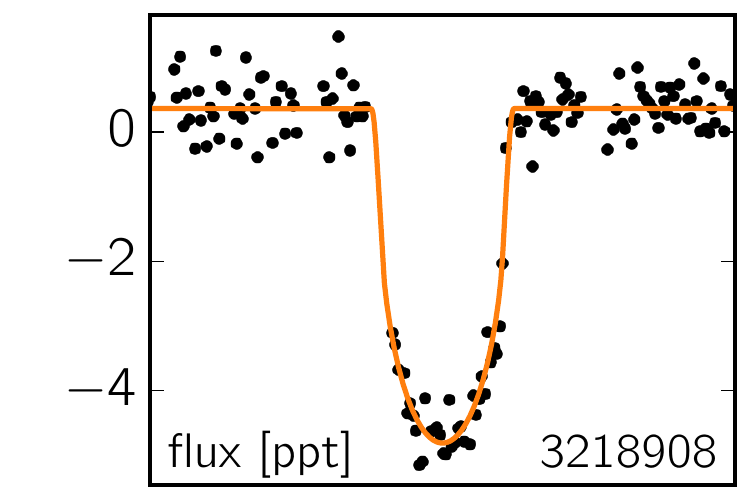}
\includegraphics[width=0.24\textwidth]{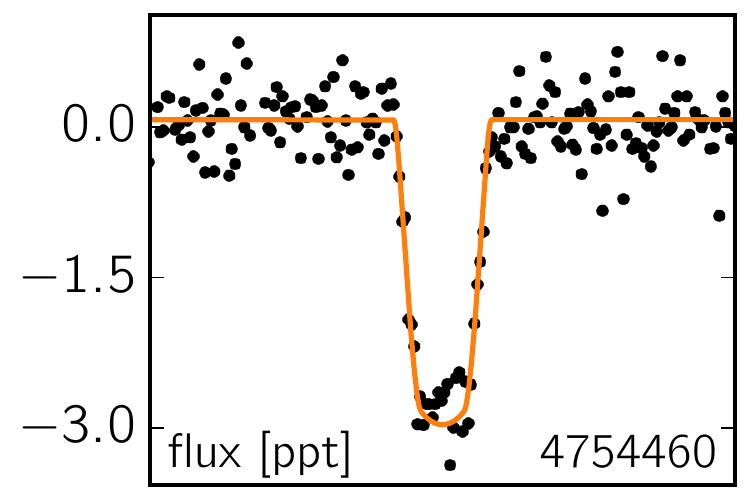}
\includegraphics[width=0.24\textwidth]{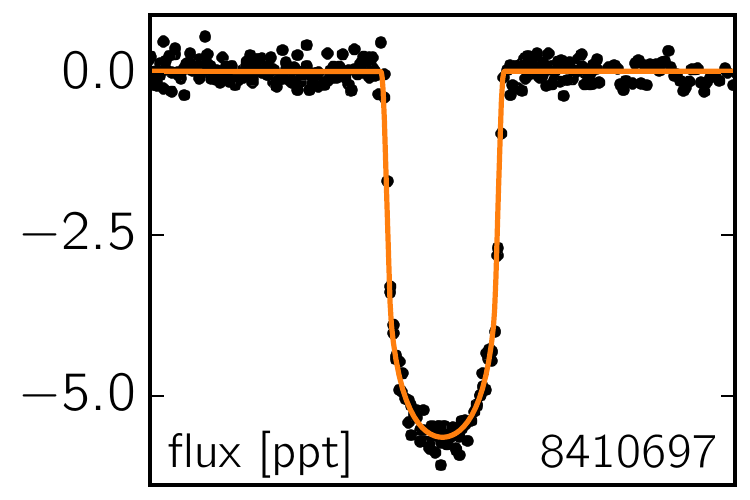}
\includegraphics[width=0.24\textwidth]{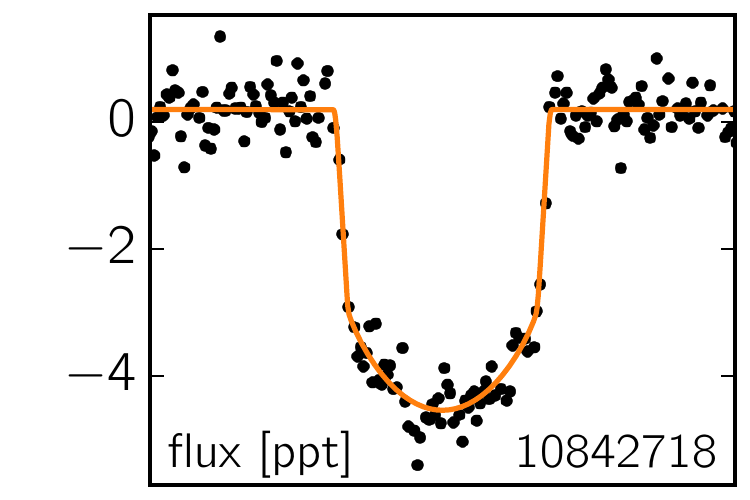}
\includegraphics[width=0.24\textwidth]{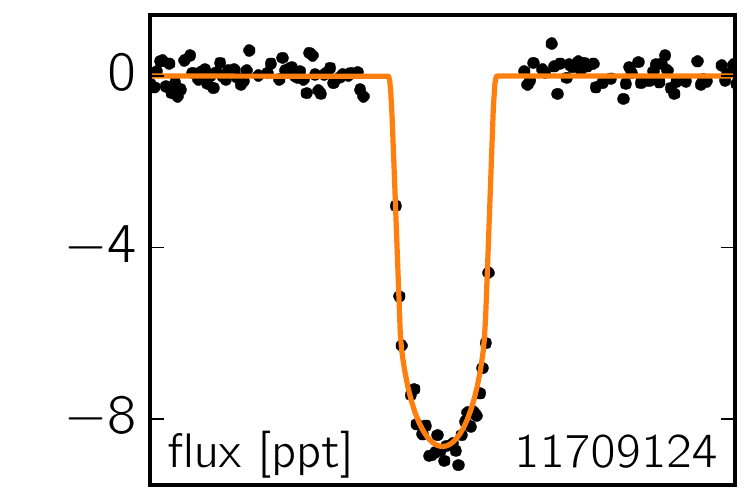}
\includegraphics[width=0.24\textwidth]{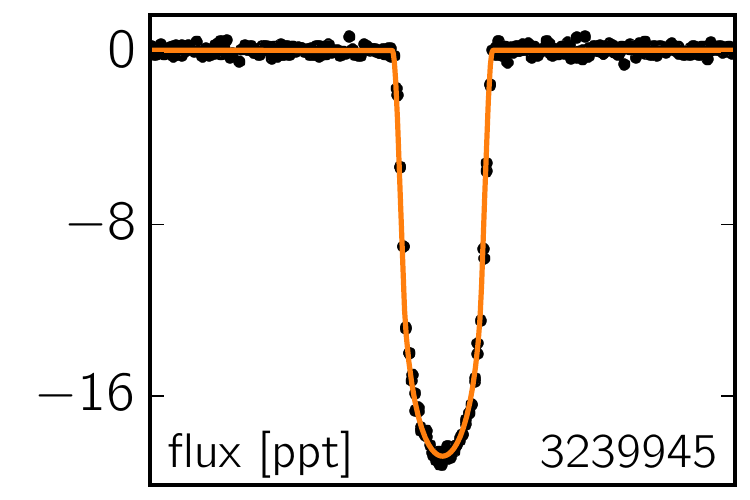}
\includegraphics[width=0.24\textwidth]{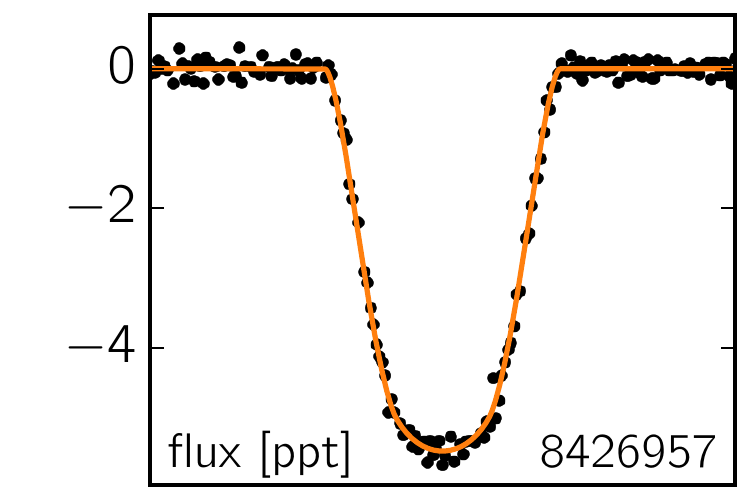}
\includegraphics[width=0.24\textwidth]{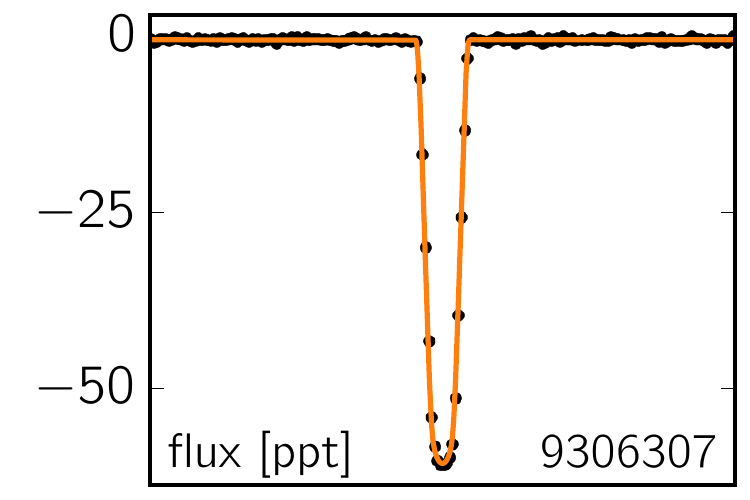}
\includegraphics[width=0.24\textwidth]{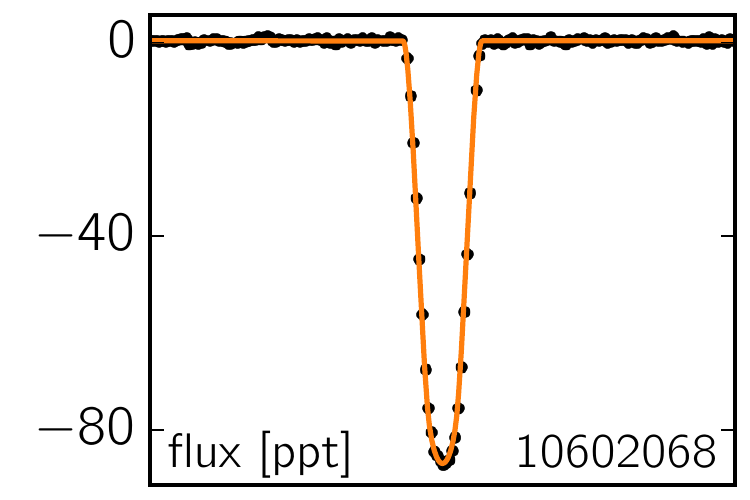}

\end{center}
\caption{%
Sections of \pdc\ light curve centered on each candidate (black) with the
posterior-median transit model over-plotted (orange).
The y-axis shows the relative apparent flux of the light curve in parts per
thousand (ppt).
Candidates with two transits are folded on the posterior-median period.
The plots are ordered by increasing planetary radius from the top-left to the
bottom-right.
\dfmfiglabel{light-curves}}
\end{figure*}

\begin{figure*}~\\
\begin{center}
\includegraphics{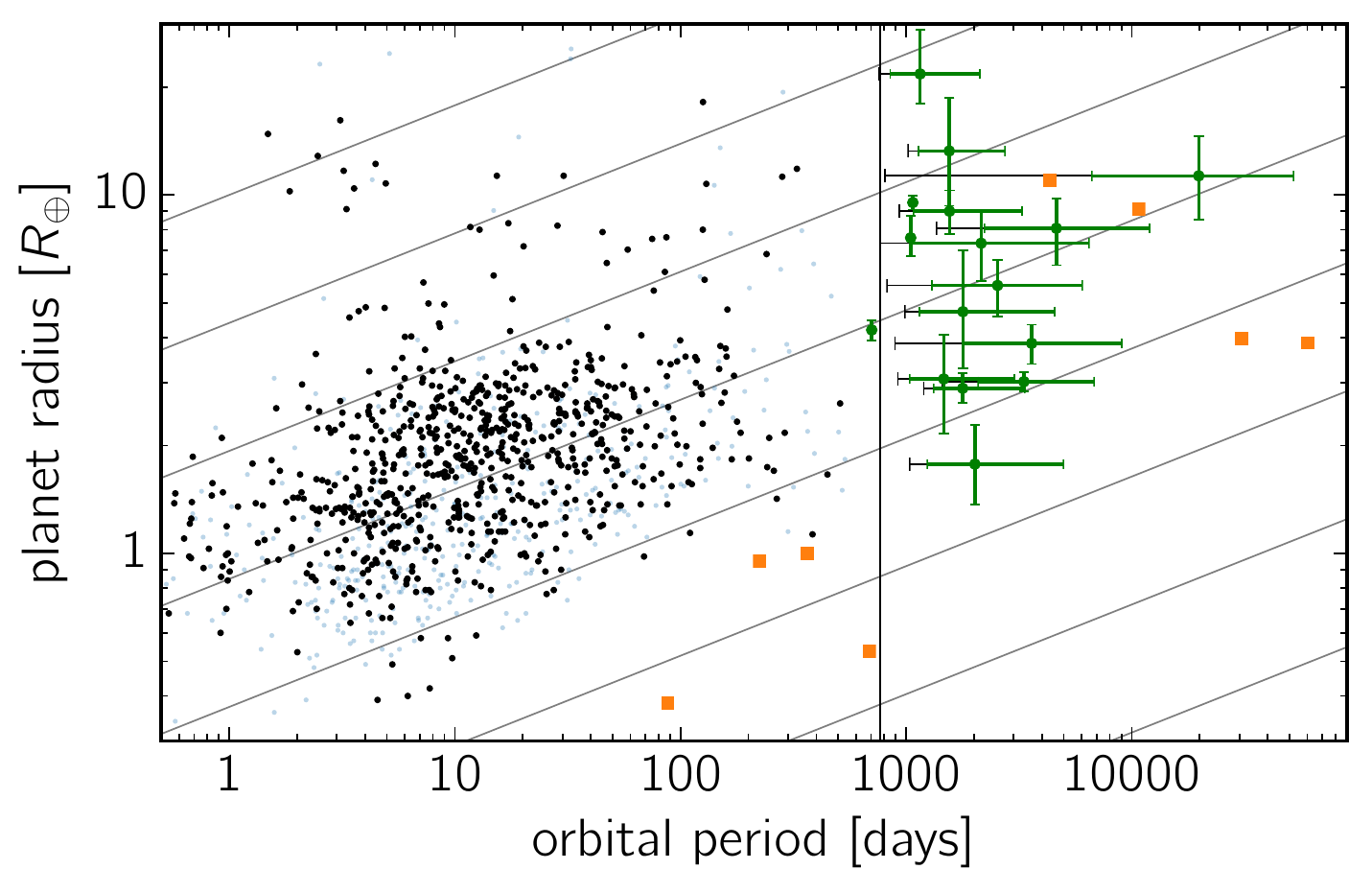}
\end{center}
\caption{%
The catalog of long-period transiting exoplanet candidates (green points with
error bars) compared to the \kepler\ candidates (blue points) and confirmed
planets \citep[black points;][]{Morton:2016} found in our target sample, and
the Solar System (orange squares).
The thin black error bars to the left of each candidate indicate the minimum
period allowed for each candidate by the prior assumption that no other
transit occurred during the baseline of \kepler\ observations of the target.
The vertical solid line shows the absolute maximum period accessible to
transit searches that require at least three transits in the \kepler\ data.
\dfmfiglabel{full-sample}}~\\
\end{figure*}

\begin{floattable}
\begin{deluxetable}{ccccl}
\tabletypesize{\footnotesize}
\caption{The signals rejected with a centroid shift or large impact parameter
\label{tab:rejects}}
\tablehead{
    \colhead{kic id} &
    \colhead{time} &
    \colhead{{d}epth} &
    \colhead{duration} &
    \colhead{reason} \\
    & \colhead{KBJD} & \colhead{ppt} & \colhead{hours} &
}
\startdata
3230491 & 315.3 & 9.0 & 7.4 & impact\\
6342758 & 553.9 & 10.3 & 9.9 & impact\\
8463272 & 641.0 & 35.5 & 4.8 & impact\\
8463272 & 1206.7 & 35.5 & 4.8 & impact\\
10668646 & 1449.2 & 5.7 & 12.4 & centroid\\
\enddata

\end{deluxetable}
\end{floattable}

\section{Empirical search completeness}\sectlabel{completeness}

To measure the completeness of the search procedure described in \sect{search}, we
exploit the fact that transit signals are sparse and rare.
Therefore, most light curves contain no transits and we can reliably measure
the recovery rate of our method on synthetic transit signals~--~with known
properties~--~injected into real light curves.
This procedure is standard practice in the transit literature and it has been
used to determine the completeness of the KOI catalog
\citep{Christiansen:2013, Christiansen:2015} and other independent transit
searches \citep{Petigura:2013, Dressing:2015, Foreman-Mackey:2015}.

To reliably capture the full structure of the search completeness function,
the simulations must sample the (high-dimensional) space of all properties
that affect the probability of detecting a transit: the stellar properties
(including variability amplitudes and time scales), the planet's physical
properties and orbital elements, and any observational effects (noise,
spacecraft pointing variations, \etc).
For the modest goals of this paper, we only need a robust constraint on the
transit detection efficiency \emph{integrated} across the target sample but,
even so, many simulations per star are required.

The procedure for measuring the recovery rate of simulated transits is as
follows:
\begin{enumerate}
{\item First, a star is randomly selected from the target list, and the \pdc\
light curve and stellar properties for that star are loaded.}
{\item Planetary properties are sampled from the distributions listed in
Table~\ref{tab:simulations} with phase uniformly distributed across the
baseline of observations. These properties are re-sampled until the transit is
visible in at least one non-flagged cadence.}
{\item The transit signal induced by this planet is computed and multiplied
into the \pdc\ light curve.}
{\item The transit search method described in \sect{search}~--~including
de-trending and \emph{all automated vetting}~--~is applied to this light
curve with the injected transit signal.}
{\item This candidate is flagged as recovered if at least one transit within
one transit duration passes all the cuts imposed by the automated vetting.}
\end{enumerate}

The fraction of recovered simulations as a function of the relevant parameters
gives an estimate of the probability of detecting an exoplanet transit with a
given set of parameters, \emph{conditioned on the fact that it transits the
star during a time when the star was being observed by Kepler}.
We will call this function $Q_{\mathrm{det},k}(\params)$ where \params\ is the
set of all parameters affecting the transit detectability and $k$ is an index
running over target stars.

\dfmfig{completeness} shows the fraction of recovered simulations as a
function of planet radius and orbital period based on \numinjs\ injected
signals.
This figure shows the transit detection efficiency falling with decreasing
planet radius.
This is the expected behavior because the depth (and signal strength) of a
transit scales with the area ratio between the planet and the star.
There is also a slight decrease in the completeness to larger planet radius.
This trend is introduced in steps~2 and~3 of the search procedure where the
tuning parameters were chosen to maximize the yield of convincing small
transit discoveries.
The decreasing completeness with orbital period is less intuitive because, on
average, the signal strength should increase as the duration of the transit
increases.
In this case, this simplistic treatment misses two important factors.
First, in step 1 of the search procedure (\sect{stepone}) only a single
transit duration is used and second, longer transits are less easily
distinguished from stellar variability and they will, therefore, be discarded
in the conservative light curve vetting step (\sect{light-curve-vetting}).

This detection efficiency must then be combined with the geometric transit
probability function and the window function.
For the star $k$, the geometric transit probability is given by
\citep{Winn:2010}
\begin{eqnarray}\eqlabel{geom-comp}
Q_{\mathrm{geom},k} (\params) &=& \frac{R_{\star,k} + R}{a_k}
    \, \frac{1 + e\,\sin\omega}{1-e^2} \\
&=& \left[\frac{4\,\pi^2}{G\,M_{\star,k}}\right]^{1/3}
    \, \left[\frac{1 + e\,\sin \omega}{1-e^2}\right]\,(R_{\star,k}+R)
    \, P^{-2/3}
\end{eqnarray}
where $R$ is the planet radius, $P$ is the orbital period, $e$ is the orbital
eccentricity, $\omega$ is the argument of periastron, $R_{\star,k}$ is
the radius of star $k$, and $M_{\star,k}$ is the star's mass.
All of these parameters are included in \params.

\response{%
In \eq{geom-comp}, the term $(R_{\star,k}+R)$ takes grazing transits into
account.
This might seem counter intuitive because, as part of the search procedure, we
rejected candidates where the maximum likelihood model had a grazing transit.
However, since the measurement of $Q_{\mathrm{det},k}$ included a cut on the
measured impact parameter, the $Q_{\mathrm{det},k}$ term already takes this
effect into account.
In other words, $Q_{\mathrm{det},k}$ quantifies the probability that a transit
of a given shape will be detected \emph{given that it transits at all} and
$Q_{\mathrm{geom},k}$~--~the way it is written in \eq{geom-comp}~--~is the
marginalized probability that the system will transit given its physical
parameters.
}

Approximating the window function using a binomial probability of observing
at least one transit, we find \citep[following][]{Burke:2014a}
\begin{eqnarray}
Q_{\mathrm{win},k} (\params) &=& \left\{\begin{array}{ll}
1 - (1 - f_{\mathrm{duty},k})^{T_k/P} & \quad\mbox{if $P \le T_k$} \\
T_k\,f_{\mathrm{duty},k} / P & \quad\mbox{otherwise}
\end{array}\right.
\end{eqnarray}
where $f_{\mathrm{duty},k}$ is the duty cycle and $T_k$ is the full
observation baseline for target $k$.

Combining these detection effects, the total detection efficiency is given by
\begin{eqnarray}
Q_k(\params) &=& Q_{\mathrm{det},k}(\params) \,
                 Q_{\mathrm{win},k} (\params) \,
                 Q_{\mathrm{geom},k} (\params) \quad.
\end{eqnarray}

So that our planet candidate catalog can be easily used for other projects, we
also provide an analytic approximation to the relevant integrated detection
efficiency function
\begin{eqnarray}
Q_\mathrm{det}(P,\,R) &=& \sum_{k=1}^{K} \int Q_\mathrm{det,k}(\params)\,
    p(\params_{\{P,\,R\}}) \dd\params_{\{P,\,R\}}
\end{eqnarray}
where $p(\params_{\{P,\,R\}})$ is the prior distribution of all the parameters
except the period and radius.
We find that a good fit to this integrated completeness is given by the
function
\begin{eqnarray}
Q_\mathrm{det}(P,\,R) &\approx&
    \frac{\mathrm{min}[\mathrm{max}[a(P)\,b(R),\,0],\,1]}
         {1+\exp\left[-k(P)\,(\ln R / R_\mathrm{J}-x(P))\right]}
\end{eqnarray}
where
\begin{eqnarray}
a(P) = a_1\,\ln P / \unit{yr} + a_2 \,,\quad
b(R) = b_1\,\ln R / R_\mathrm{J} + b_2 \,,\quad \\
k(P) = k_1\,\ln P / \unit{yr} + k_2 \,,\,\mathrm{and}\quad
x(P) = x_1\,\ln P / \unit{yr} + x_2 \,.
\end{eqnarray}
When fit to the set of \numinjs\ injected transits, the best fit parameters
are given in Table~\ref{tab:completeness} and the approximation is plotted
in \dfmfig{completeness-analytic}.
Note that we do not use this approximation in the following analysis but
instead compute the relevant integrals using the injection results directly.

\begin{floattable}
\begin{deluxetable}{lc}
\tabletypesize{\footnotesize}
\caption{Distributions of physical parameters for transit simulations
\label{tab:simulations}}

\tablehead{%
    \colhead{name} & \colhead{distribution}
}
\startdata
period & $\log P \sim \mathcal{U}(\log 2\unit{yr},\,\log 25\unit{yr})$ \\
radius ratio & $\log R_\mathrm{P}/R_\star \sim
    \mathcal{U}(\log 0.02,\,\log 0.2)$ \\
impact parameter & $b \sim \mathcal{U}(0,\,1+R_\mathrm{P}/R_\star)$ \\
\multirow{2}{*}{eccentricity} & $e \sim \beta(1.12,\,3.09)$\tablenotemark{a} \\
    & $\omega \sim \mathcal{U}(-\pi,\,\pi)$\\
\multirow{2}{*}{limb darkening} & $q_1 \sim \mathcal{U}(0,\,1)$ \\
                                & $q_2 \sim \mathcal{U}(0,\,1)$ \\
\enddata

\tablenotetext{a}{\citet{Kipping:2013a}}
\end{deluxetable}
\end{floattable}

\begin{figure*}[p]~\\
\begin{center}
\includegraphics[width=\textwidth]{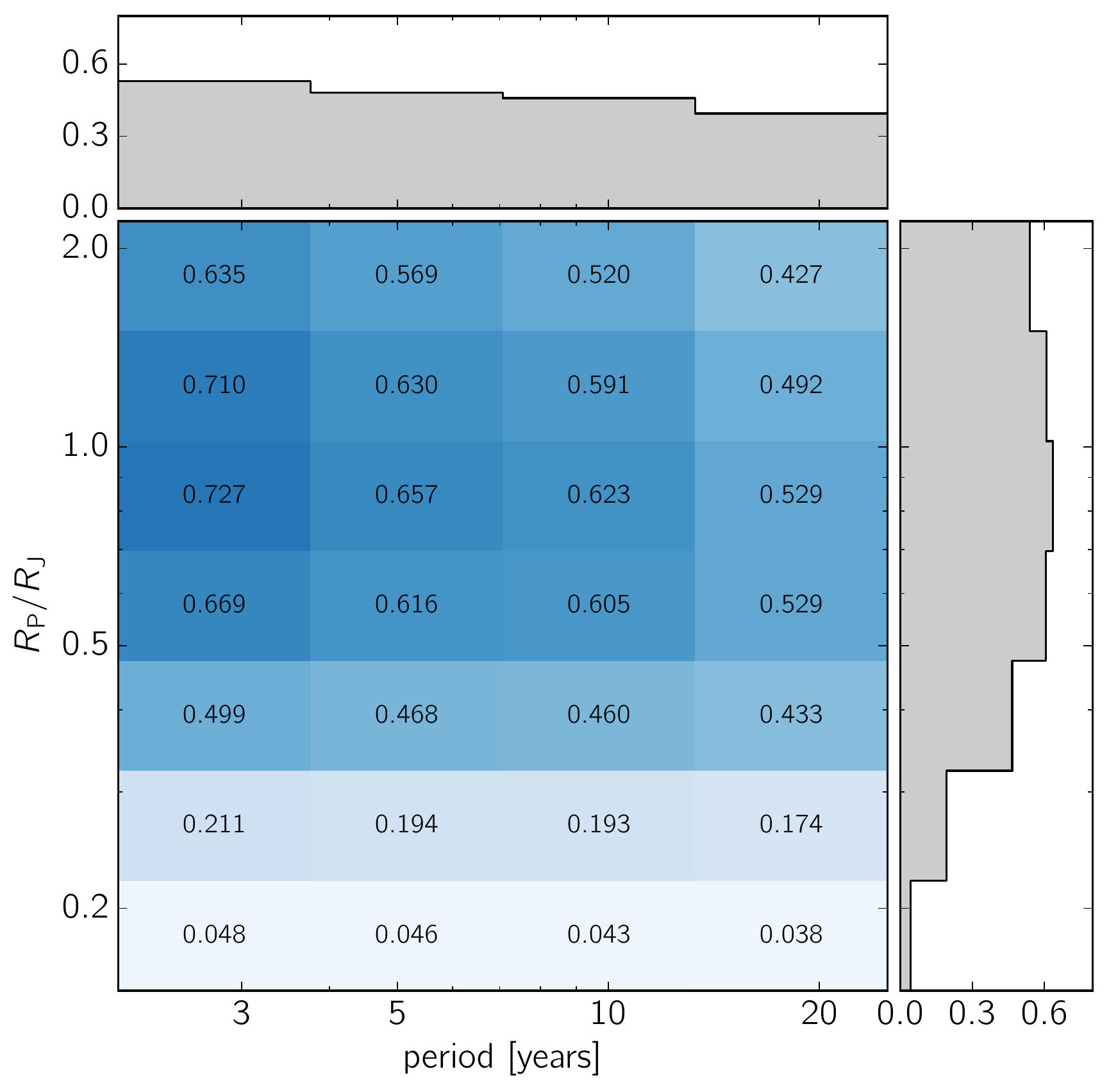}
\end{center}
\caption{%
An empirical estimate of the search completeness as a function of planet
radius and orbital period.
In each bin, the completeness is estimated by the fraction of recovered
simulations.
The projected histograms show the integrated completeness as independent
functions of period and radius.
\dfmfiglabel{completeness}}
\end{figure*}

\begin{figure*}[htbp]~\\
\begin{center}
\includegraphics[width=0.6\textwidth]{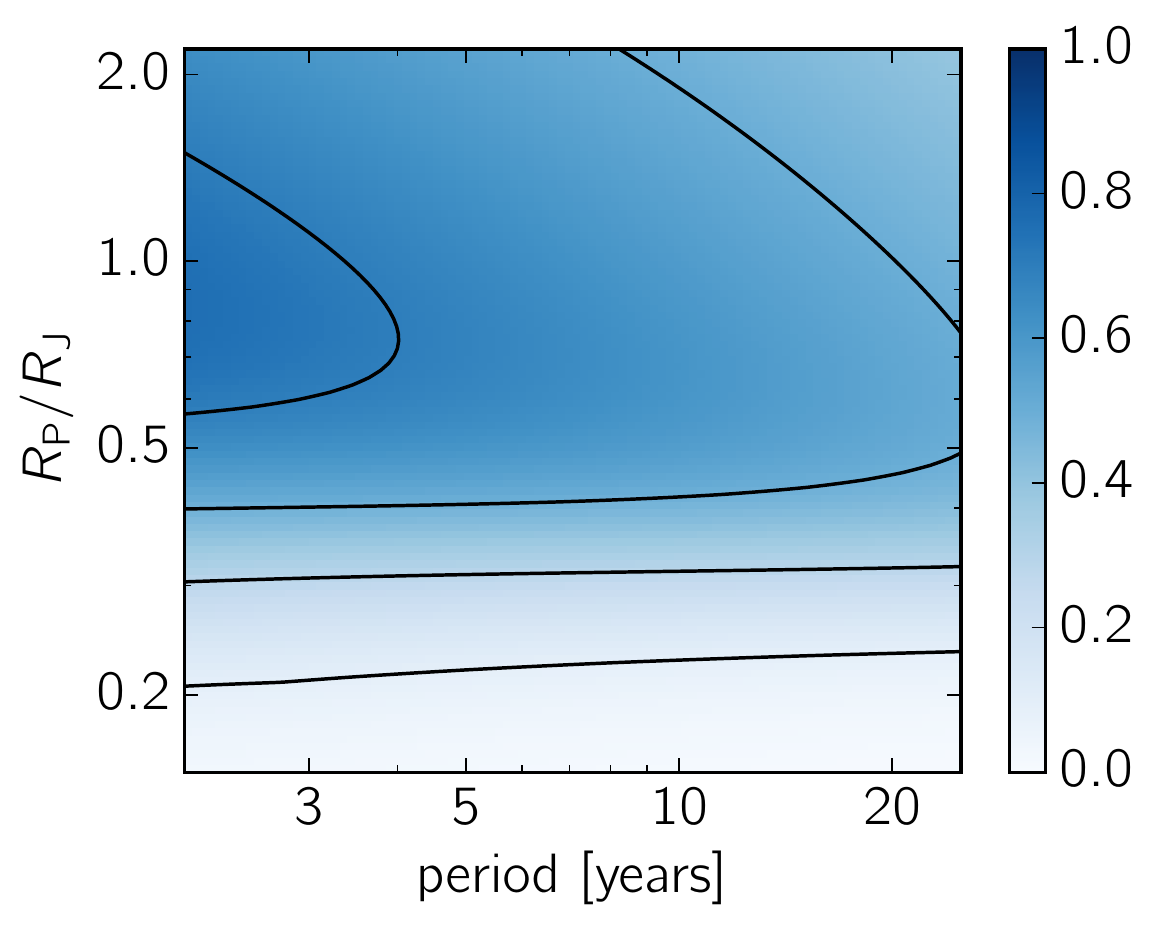}
\end{center}
\caption{%
An analytic approximation to \dfmfig{completeness} with the same color scale.
The contours indicate the 0.1, 0.3, 0.5, and 0.7 levels.
\dfmfiglabel{completeness-analytic}}
\end{figure*}

\begin{floattable}
\begin{deluxetable}{cc|cc}
\tabletypesize{\footnotesize}
\caption{The fit parameters for the analytic approximation to the completeness
function
\label{tab:completeness}}
\tablehead{%
    \colhead{parameter} & \colhead{value} &
    \colhead{parameter} & \colhead{value}
}
\startdata
$a_1$ & $\paramc$ & $k_1$ & $\parame$ \\
$a_2$ & $\paramd$ & $k_2$ & $\paramf$ \\
$b_1$ & $\parama$ & $x_1$ & $\paramg$ \\
$b_2$ & $\paramb$ & $x_2$ & $\paramh$ \\
\enddata
\end{deluxetable}
\end{floattable}

\section{The occurrence rate of long-period exoplanets}\sectlabel{occurrence}

Using the catalog of exoplanet discoveries (\sect{catalog}) and the
measurement of the search completeness (\sect{completeness}), we can now
estimate the occurrence rate of long-period exoplanets.
To simplify the analysis, we will make the strong assumption that none of the
candidates are astrophysical false positives (the eclipse or occultation
of a stellar mass companion, around either the target star or a
faint background star).
We revisit this assumption and discuss its validity in the following section.
As a further simplification, we also neglect the measurement uncertainties on
the planet parameters (including orbital period).
This assumption is justified because we are only making high-level
measurements of the mean occurrence rate in bins larger than the
uncertainties.

Assuming a Poisson likelihood, the occurrence rate density in a volume
$V$~--~defined as $P_\mathrm{min} \le P < P_\mathrm{max}$ and
$R_\mathrm{min} \le R < R_\mathrm{max}$~--~is
\citep[see, for example, the Appendix of][]{Foreman-Mackey:2014}
\begin{eqnarray}\eqlabel{rate-density}
\Gamma_V \equiv \frac{\dd^2 N}{\dd\ln P\dd\ln R} &=&
    \frac{C(P_\mathrm{min},\,P_\mathrm{max};\,R_\mathrm{min},\,R_\mathrm{max})}
         {Z(P_\mathrm{min},\,P_\mathrm{max};\,R_\mathrm{min},\,R_\mathrm{max})}
\end{eqnarray}
where $N$ is the expected number of planets per G/K dwarf, $C(\cdots)$ is the
number of detected planets in the volume, and
\begin{eqnarray}
Z(P_\mathrm{min},\,P_\mathrm{max};\,R_\mathrm{min},\,R_\mathrm{max}) &=&
    \sum_{k=1}^{K} \int p(\params_{\{P,\,R\}})\,
    Q_k(\params)\,\bvec{1}[P,\,R\in V]\dd \params
\end{eqnarray}
where $p(\params_{\{P,\,R\}})$ is the prior distribution of all the parameters
except the period and radius and $\bvec{1}[\cdot]$ is 1 if the argument is
satisfied and 0 otherwise.
Using the $J$ injections sampled uniformly in period and radius and other
parameters from $p(w_{\{P,\,R\}})$,
\begin{eqnarray}\eqlabel{approxnorm}
Z(P_\mathrm{min},\,P_\mathrm{max};\,R_\mathrm{min},\,R_\mathrm{max}) &\approx&
    \frac{K\,V}{J}\,\sum_{j=1}^{J} Q_{k_j}(w^{(j)})
\end{eqnarray}
where the sum is over all injections in the volume $V$.

Using the injection results from \sect{completeness} and the catalog of
discoveries from \sect{catalog}, we compute the occurrence rate in the period
range 2 to 25 years and in two radius bins between 0.1 and
$1.0\,R_\mathrm{J}$.
The calculated occurrence rates are listed in Table~\ref{tab:occurrence}.
Integrating the two bin model in this range, we find an expected occurrence
rate of
\begin{eqnarray}
N_{0.1\,R_\mathrm{J}-1\,R_\mathrm{J},\,2\unit{yr}-25\unit{yr}} &=&
    \intocc \unit{planets}
\end{eqnarray}
per G/K dwarf with radii in the range $0.1\,R_\mathrm{J}-1\,R_\mathrm{J}$ and
periods in the range $2\unit{yr}-25\unit{yr}$.
This result is qualitatively consistent with the Solar System where there is
one planet~--~Jupiter~--~in this parameter range and Saturn is just outside
the range with an orbital period of 29~years.
In \sect{comparison}, we compare with similar occurrence rate estimates from
the literature.

The occurrence rates given here should be interpreted with a few caveats in
mind.
First, when we inferred the periods of the planets with only one transit, we
assumed that the period was long enough that no other transit occurred during
the \kepler\ lifetime.
This neglects the small but non-negligible posterior probability~--~less than
one percent for the typical candidate~--~that a second transit might have
occurred in a data gap.
All of the candidates in our catalog are consistent with having periods this
long but the geometric transit probability decreases quickly with orbital
period.
For the purposes of this \paper, we neglect this effect because its rigorous
treatment is subtle, but comment that this would only ever decrease the
occurrence rate estimate.   Second, we assume that each planet candidate transits
the star that is characterized by \citet{Huber:2014}; we assert that each planet
does not transit a fainter companion star or a background star.  If the planet
does transit a companion star, then the companion star must be fainter, and hence
denser, causing the period to be underestimated.  If the planet transits
a background star, it is more likely to be a giant star due to Malmquist bias,
hence the density of the star and period of the planet would be overestimated.
Either of these scenarios has a small
probability, so we expect that our population estimates will stand, while
the parameter estimates for individual candidates should be taken as provisional
until more detailed follow-up is carried out, including high-contrast imaging,
high-resolution spectroscopy, and parallax measurements.  Third, we assume
that the \citet{Huber:2014} parameters are accurate for each star that is transited by
a planet candidate, and that each transit is unaffected by blending.
Malmquist bias, Eddington bias, and metallicity bias may affect the stellar parameters
\citep{Gaidos:2012}, and so we again caution that the individual parameter
estimates should be taken as provisional until more detailed follow-up is
completed.

\begin{floattable}
\begin{deluxetable}{cccc}
\tabletypesize{\footnotesize}
\caption{The occurrence rate density in two radius bins \label{tab:occurrence}}
\tablehead{
    \colhead{$R_\mathrm{min}\,[R_\mathrm{J}]$} &
    \colhead{$R_\mathrm{max}\,[R_\mathrm{J}]$} &
    \colhead{rate density\tablenotemark{a}} &
    \colhead{integrated rate\tablenotemark{b}}
}
\startdata
$0.1$ & $0.4$ & $0.45\pm0.20\,(0.36\pm0.16)$ & $1.57\pm0.70\,(1.26\pm0.56)$ \\
$0.4$ & $1.0$ & $0.18\pm0.07\,(0.16\pm0.06)$ & $0.42\pm0.16\,(0.36\pm0.14)$ \\
$0.1$ & $1.0$ & $0.24\pm0.07\,(0.22\pm0.06)$ & $1.41\pm0.41\,(1.29\pm0.37)$ \\
\enddata

\tablenotetext{a}{The rate density is given by \eq{rate-density} and the
value in parentheses is computed assuming one candidate is a false positive
(\eqalt{rate-minus}).}
\tablenotetext{b}{The integrated rate is computed by integrating the rate
density over the bin. Note that the first two rows do not sum to the last row
because each row is computed assuming that the rate density is uniform
across the bin.}
\tablecomments{These values are computed in the period range 2--25~years.}
\end{deluxetable}
\end{floattable}

\section{Astrophysical false positives}\sectlabel{false-positives}

Various configurations of eclipsing binary stars can mimic the signal of a
transiting planet.  However, the occurrence rate calculation
presented in the preceding section assumes no astrophysical false positives
among the candidates identified in this work.  In this Section we explore the
validity of this assumption.

While an eclipsing binary (EB) typically produces a photometric dip much
deeper than a transiting planet, the depth of the signal may be comparable to
that of a planet if the eclipse is grazing, or if the EB only comprises a small
fraction of the total light in the photometric aperture~--~a so-called blended
eclipsing binary (BEB).  Additionally, if a binary star has an eccentric
orbit, it may be oriented so as to present only a secondary occultation and
not a primary eclipse, causing a shallow and potentially flat-bottomed
photometric dip without an accompanying tell-tale deep primary signal.

To determine to what extent the catalog of detections presented in this work
may contain such false positives, we simulate populations of detected signals
to predict how many we should expect.  To accomplish this, we use the Python
package \exosyspop\ \citep{Morton:2016a}, which
we developed for this purpose and utilizes the \project{isochrones}, \project{vespa},
and \project{batman} packages \citep{Morton:2015, Morton:2015b, Kreidberg:2015}
 for simulations of stellar populations and their eclipses.

With \exosyspop\, one can define the parameters of a population model and
generate synthetic catalogs according to the model (and the parameters of a
survey) very efficiently.  For example, a population of EBs may be defined by
a binary fraction, power-law distributions in mass ratio and period (within
given bounds), and a beta distribution for eccentricity.  This population,
initialized with a catalog of target stars (each of which has a duty cycle and
total span of observation), may then be ``observed,'' returning a catalog of
objects detectable via either primary or secondary eclipse (according to
randomly oriented orbital geometries and accounting for observation duty cycle
and data span).  This synthetic catalog includes signal-to-noise estimates of
both the primary and secondary eclipses, the number of detected primary and
secondary eclipses, and the trapezoidal shape parameters of each detection
(depth, duration, and ingress-to-duration ratio, as defined in
\citealt{Morton:2012}).

In order to predict how many EBs or BEBs we might expect to detect in this
particular search of \kepler\ data, we first need to choose reasonable parameters
for the binary star population.  To do this, we calibrate the population
parameters using the catalog of detected \kepler\ eclipsing binaries.  We
find that a binary fraction of 25\% between periods of 20 days and 25 years with
a log-flat period distribution and eccentrities distributed according
to $\beta(0.8, 2.0)$ is able to reproduce well both the number and period
distribution of observed \kepler\ EBs between 20 and 1000 days.  We thus fix these
binary star population parameters for our subsequent EB and BEB simulations.

To simulate synthetic populations of EB detections, we assign binary stars  to
the \kepler\ target list described in \sect{target-selection} according to the
above EB population parameters.  We consider an EB to be detected if it
presents fewer than three eclipses (either primary or secondary, but not
both), if the signal-to-noise ratio is $>$15, and the duration of the
detected eclipse is $<$2.5\,d.  In 100 realizations of these synthetic
observations, we see $7.2 \pm 2.5$ single- or double-eclipsing EB signals.

To simulate BEBs, we assume an exponentially varying background field star
density across the \kepler\ field, from 0.005\,arcsec$^{-2}$ at a Galactic
latitude $b=20$ to 0.05\,arcsec$^{-2}$ at $b=5$ (matching up well with the
simulations of \citet{Morton:2011} at many different Galactic latitudes). Each
\kepler\ target star is then assigned a number of background stars drawn from
a Poisson distribution with mean given by the expected number of stars to be
found within a circle of 4\,arcsec radius, given the  appropriate density at
its Galactic lattitude.   We draw the specific background stars from a
TRILEGAL \citep{Girardi:2005}  field star simulation toward the center of the
\kepler\ field.  Binary companions are then assigned to these background stars
according to the same stellar binary population distribution as the EB
population above, and synthetic detected populations  are ``observed''
according to the same rules (accounting appropriately for the diluted eclipse
depths in the \kepler\ bandpass).  In 100 synthetic observations, we see an
average of only 0.41 detected BEBs.

These prediction results suggest that we should indeed expect to see some
astrophysical false positives in our search.  However, this does not mean that  we
should fear that $\sim$7 of the planet candidates might be EBs. In particular,
we note that these simulations do not include the full vetting procedure
described in \sect{catalog}, and it is likely that the three impact-parameter-rejected
candidates are EBs and that the centroid-rejected candidate is a BEB.  Thus,
we might expect maybe two or three additional false positives among our planet
candidates.

In order to more precisely quantify which of the candidates might indeed be false
positives, we can inspect the synthetic observation simulations in more detail. In
particular, we can analyze the shape distribution of the different scenarios and
compare them with the observed shapes of the actual \kepler\ detections in order
to quantify the probability that each of them may be a false positive.
These distributions and the observed shape parameters are plotted in
\dfmfig{shape-kdes}.

Following the method of \citet{Morton:2012} used to compute false positive
probabilities for the regular \kepler\ KOI catalogs \citep{Morton:2016}, we
can calculate the posterior probability for each of our candidates to belong
to each of the three scenarios we consider (EB, BEB, or planet) as follows:
\begin{eqnarray}
{\rm Pr}_i = \frac{\pi_i \mathcal L_i}{\sum_{j} \pi_j \mathcal L_j},
\end{eqnarray}
where the $\pi$ factors are the hypothesis priors, $\mathcal L$ are
the hypothesis likelihoods, and the sum over $j$ is over all the hypotheses.
In this case, we determine the relative hypothesis priors from the synthetic
observations, using the mean numbers of ``observed'' EBs (7.2) and BEBs (0.4),
and choosing the expected number of planets to be 12.  We calculate the hypothesis
likelihoods using the depth/duration distributions of synthetic populations
of each scenario and evaluating these distributions at the observed depths and
durations of each candidate signal.  To estimate the expected shape distribution
of the planet scenario, we define a custom \exosyspop\ population of planets
according to the two-bin population model described by the
median posterior values in Table~\ref{tab:occurrence} and
generate a population of 1000 detected signals.

\response{We list the probability that each candidate is a planet in
Table~\ref{tab:catalog}.} We find that the two deepest signals in our
candidate catalog (9306307 and  10602068) are very likely to be EBs, though we
note that this result may be dominated by the fact that our planet population
is fixed to have a maximum radius of 1\,$R_{\rm J}$.  Most of the rest of the
candidates  have false positive probabilities below 10\%.  We do note that as
discussed in \sect{catalog}, 4754460 (for which we calculate a 5\% false positive
probability) does show a partial deep eclipse right at the end of the \kepler\
data that indicates that it is most likely an EB.  Apart from this, the expected
number of false positives among the candidates with $R < R_{\rm J}$, according
to these calculations, is about one.

In light of this result, we demonstrate the sensitivity of our measured
occurrence rates on contamination by computing a second constraint on
$\Gamma_V$ for each volume with one candidate removed.
In this case, \eq{rate-density} would be replaced by
\begin{eqnarray}\eqlabel{rate-minus}
\tilde{\Gamma}_V &\equiv&
    \frac{C(P_\mathrm{min},\,P_\mathrm{max};\,R_\mathrm{min},\,R_\mathrm{max})
          - 1}
         {Z(P_\mathrm{min},\,P_\mathrm{max};\,R_\mathrm{min},\,R_\mathrm{max})}
\quad.
\end{eqnarray}
These updated rates are listed in Table~\ref{tab:occurrence}.
In each case, the results are consistent within the uncertainties but the
difference can be used to get a qualitative sense of the systematic
uncertainty introduced by the false positive population.

\response{%
We note that in the above procedure we have not corrected our predictions
for the fact that our search has explicitly excluded KOIs
that host known \kepler\ EBs---if any of these excluded systems show fewer than
three eclipses and do not present both primary and secondary eclipses, then they
should also should perhaps count towards the number of EBs we should have
expected to find in this survey.  However, as the \kepler\ EB catalog does not provide information
on whether both primary and secondary eclipses are detected, we neglect this correction.
We note that this is a conservative decision, in the sense that accounting for the
effect of excluded EBs on our predictions would only further decrease the FPP of
the planetary signals, as they would be even less likely to be caused by EBs.
}

\begin{figure*}[htbp]~\\
\begin{center}
\includegraphics[width=\textwidth]{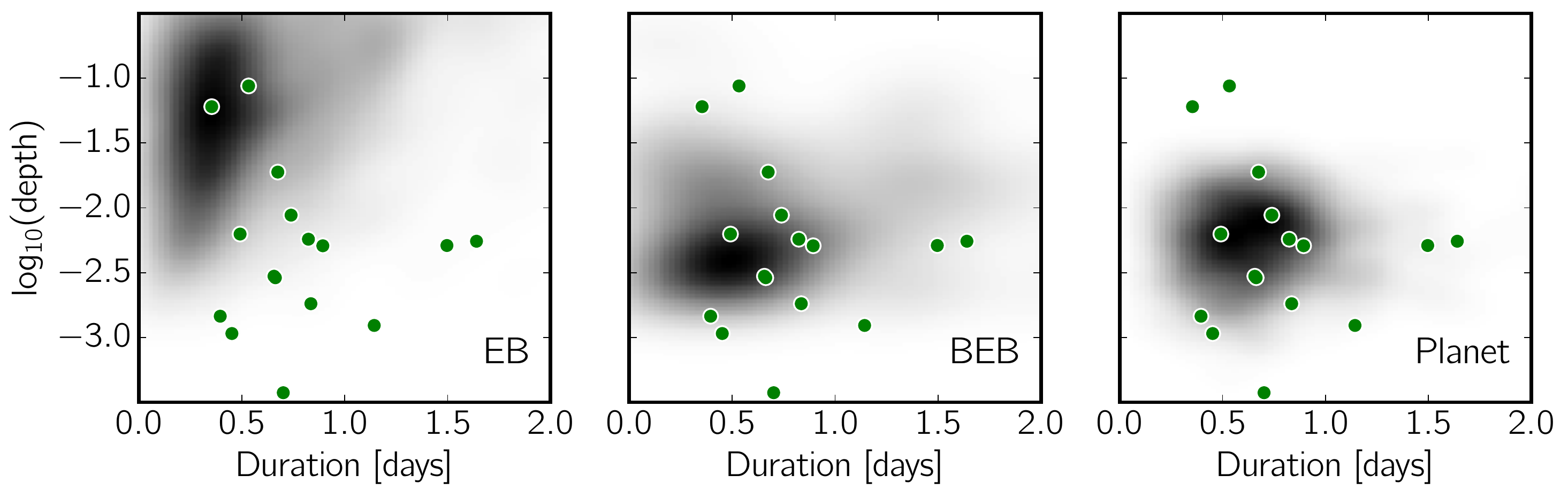}
\end{center}
\caption{%
Predicted eclipse shape distributions for the two false positive scenerios and
exoplanet transits (grayscale heatmap).
In this figure, the relative normalization of the maps are arbitrary but the
absolute normalization is discussed in \sect{false-positives}.
The green points show the shape parameters of the long-period exoplanet
candidates from Table~\ref{tab:catalog}.
\dfmfiglabel{shape-kdes}}
\end{figure*}

\section{Comparison with the literature}\sectlabel{comparison}

The population of long-period planets has been previously studied using radial
velocity, microlensing, and direct imaging surveys.
These methods all measure the occurrence rate as a function of planet mass
instead of radius.
Using the transit method, however, we do not directly measure the mass of the
planet.
Therefore, to compare our results with the literature, we must rely on a
mass--radius (M--R) relationship constructed using exoplanets with both mass
and radius measurements \citep[for example][]{Weiss:2014, Wolfgang:2016,
Chen:2016} to predict the expected masses of the transiting planets.

Table~\ref{tab:masses} lists gives constraints on the predicted masses of the
exoplanet candidates using the probabilistic M--R relationship from
\citet{Chen:2016} and taking uncertainties in the planet radius and
statistical uncertainties in the M--R parameters.
We compare the predictions with the predictions from \citet{Wolfgang:2016} and
find similar values with smaller uncertainties and choose to use the
\citet{Chen:2016} relationship because it is more conservative in the relevant
range of parameter space.

A detailed discussion of the systematic effects introduced by the use of a
M--R relationship is beyond the scope of this \paper\ but it is worth
noting that all published relationships are based on exoplanets much closer to
their host star than any of the candidates discussed here.
This effect would cause the masses of these cool planets to be systematically
underestimated.

Using the same M--R relationship, we also compute the completeness of our
transit search as a function of planet mass and orbital semi-major axis.
This function is plotted in \dfmfig{completeness-am} with the same color
scheme as \dfmfig{completeness}.
These injections and the predicted masses and measured semi-major axes of the
candidates can then be used to estimate the occurrence rate in
mass--semi-major axis units using the method from \sect{occurrence}.
One small change to \eq{approxnorm} is necessary to account for the fact that
the injections were not made uniformly in $\ln M$ and $\ln a$.
We numerically estimate the prior distribution in mass and semi-major axis
from which the injections were drawn $\tilde{p}(\ln a,\,\ln M)$ and
\eq{approxnorm} becomes
\begin{eqnarray}
Z(a_\mathrm{min},\,a_\mathrm{max};\,M_\mathrm{min},\,M_\mathrm{max}) &\approx&
    \frac{K\,V}{J}\,\sum_{j=1}^{J}
    \frac{Q_k(w^{(j)})}{\tilde{p}(\ln a^{(j)},\,\ln M^{(j)})} \quad.
\end{eqnarray}
Using this result, we find that the mean occurrence rate density in the range
$0.01\,M_\mathrm{J} \le M < 20\,M_\mathrm{J}$ and $1.5\unit{au} \le a <
9\unit{au}$ is
\begin{eqnarray}\eqlabel{massocc}
\frac{\dd^2 N}{\dd\ln M\,\dd\ln a} &=& \masssemiocc
\end{eqnarray}
where $N$ is the expected number of planets per G/K dwarf.
This result and the equivalent result as a function of planet mass and orbital
period are listed in Table~\ref{tab:occrates}.

The uncertainty in \eq{massocc} and Table~\ref{tab:occrates} does not take
into account the uncertainties in the mass estimates or any systematic noise
in the mass--radius relationship.
Therefore, these specific results should be taken with the appropriate grain
of salt but predictions in these parameter spaces ease comparison with
occurrence rates measured computed using different methods.

\response{%
\citet{Clanton:2016} studied the occurrence rate of long-period giant planets
orbiting M dwarfs by combining results from radial velocity, microlensing, and
direct imaging surveys.
In the period and mass range $10^3-10^4\unit{days}$ and $10-10^4\,M_\oplus$,
they find a mean occurrence rate density of
\begin{eqnarray}
\frac{\dd^2 N}{\dd\ln M\,\dd\ln P} &=& 0.023
\end{eqnarray}
per M dwarf with large uncertainty.}
This result is slightly lower than our estimated rate for a similar range of
masses and periods but around G/K dwarfs.
\response{%
This difference is consistent with previous observational and theoretical
results that cooler stars host fewer long-period giant planets \citep[for
example][]{Laughlin:2004, Cumming:2008, Clanton:2016}.
}

Recently, \citet{Bryan:2016} studied the frequency of long-period giant
planets in systems with inner hot Jupiters based on long-baseline radial
velocity monitoring of these systems \citep{Knutson:2014}.
In this sample, the computed occurrence rate of long-period giant planets was
found to be
\begin{eqnarray}
\frac{\dd^2 N}{\dd\ln M\,\dd\ln a} &=& 0.125 \pm 0.012
\end{eqnarray}
in the range $1-20\,M_\mathrm{J}$ and $5-20\unit{au}$.
This result is about a factor of two larger than our estimate
(\eqalt{massocc}) once again suggesting that cold Jupiters might
preferentially occur in systems with inner planets~--~or that the
presence of cold Jupiters encourages the formation of hot Jupiters.

A recent review of the occurrence rate estimates based on direct imaging
surveys \citep{Bowler:2016} reports the upper limit on the occurrence rate of
giant planets orbiting F/G/K dwarfs as $<6.8\%$ in the range
$5-13\,M_\mathrm{J}$ and $10-100\unit{au}$.
Converted to a rate density, this gives
\begin{eqnarray}
\frac{\dd^2 N}{\dd\ln M\,\dd\ln a} &<& 0.03 \quad.
\end{eqnarray} This value is lower than the value computed using our sample in
Table~\ref{tab:occrates} but this is consistent with the fact that direct
imaging is sensitive to the potentially less common large planets at wider
separations than detections with the transit method.

As a final comparison, we repeated the analysis of \citet{Burke:2015} and fit
a double power-law occurrence rate to the short-period \kepler\ planet
candidates\footnote{The analysis was adapted from publicly available code that
was demonstrated to reproduce the same results as \citet{Burke:2015} by
\citet{Foreman-Mackey:2015a}.} and extrapolated to the center of the two bins
where we computed the occurrence rate.
At a period of 7~years and a radius of $0.2\,R_\mathrm{J}$, the extrapolated
occurrence rate density is $0.73\pm0.28$ and at a radius of
$0.6\,R_\mathrm{J}$, the extrapolated rate density is $0.15\pm0.05$.
These extrapolated values are qualitatively consistent with the rates listed
in Table~\ref{tab:occurrence} but we note that extrapolations and their
statistical uncertainties should not be taken too seriously.

\begin{floattable}
\begin{deluxetable}{ccccccccc}
\tabletypesize{\scriptsize}
\caption{The predicted masses and radial velocity semi-amplitudes for the
candidates from Table~\ref{tab:catalog} \label{tab:masses}}
\tablehead{
    \colhead{kic id} &
    \colhead{Kp} &
    \colhead{radius} &
    \colhead{mass} &
    \colhead{period} &
    \colhead{$t_0$} &
    \colhead{semi-major axis} &
    \colhead{semi-amplitude} &
    \colhead{$K / P$} \\
    && \colhead{$R_\mathrm{J}$} & \colhead{$M_\mathrm{J}$} & \colhead{year} &
    \colhead{KBJD} & \colhead{au} & \colhead{m s$^{-1}$} &
    \colhead{m s$^{-1}$ yr$^{-1}$}
}
\rotate
\startdata
3218908 & $14.6$ & $0.514_{-0.093}^{+0.092}$ & $0.079_{-0.038}^{+0.074}$ & $7.0_{-3.4}^{+9.5}$ & $766.6722_{-0.0114}^{+0.0096}$ & $3.4_{-1.2}^{+2.7}$ & $1.41_{-0.77}^{+1.57}$ & $0.20_{-0.15}^{+0.49}$\\
3239945 & $14.0$ & $0.876_{-0.039}^{+0.039}$ & $6.5_{-6.3}^{+38.1}$ & $2.9328721_{-0.0000026}^{+0.0000026}$ & $420.28714_{-0.00068}^{+0.00069}$ & $1.864_{-0.023}^{+0.025}$ & $157.5_{-153.0}^{+900.5}$ & $53.7_{-52.2}^{+307.1}$\\
4754460 & $14.9$ & $0.67_{-0.15}^{+0.16}$ & $0.140_{-0.078}^{+9.289}$ & $5.9_{-3.0}^{+11.8}$ & $826.8369_{-0.0046}^{+0.0046}$ & $3.1_{-1.2}^{+3.4}$ & $2.5_{-1.5}^{+185.3}$ & $0.46_{-0.39}^{+35.87}$\\
6551440 & $13.6$ & $0.282_{-0.083}^{+0.093}$ & $0.028_{-0.015}^{+0.033}$ & $4.0_{-1.2}^{+4.2}$ & $1039.0589_{-0.0037}^{+0.0037}$ & $2.50_{-0.52}^{+1.55}$ & $0.57_{-0.33}^{+0.77}$ & $0.133_{-0.094}^{+0.230}$\\
8410697 & $13.4$ & $0.698_{-0.078}^{+0.107}$ & $0.157_{-0.081}^{+11.382}$ & $2.8688097_{-0.0000054}^{+0.0000053}$ & $542.1231_{-0.0013}^{+0.0013}$ & $1.925_{-0.038}^{+0.054}$ & $3.6_{-1.9}^{+262.6}$ & $1.26_{-0.65}^{+91.55}$\\
8426957 & $13.6$ & $1.04_{-0.25}^{+0.30}$ & $3.8_{-3.6}^{+47.5}$ & $54.2_{-36.1}^{+88.4}$ & $784.677_{-0.013}^{+0.013}$ & $14.1_{-7.4}^{+13.0}$ & $33.8_{-32.4}^{+389.5}$ & $0.73_{-0.71}^{+10.01}$\\
8505215 & $13.0$ & $0.277_{-0.017}^{+0.017}$ & $0.028_{-0.012}^{+0.022}$ & $9.1_{-3.4}^{+9.5}$ & $140.0492_{-0.0018}^{+0.0017}$ & $4.0_{-1.1}^{+2.5}$ & $0.45_{-0.21}^{+0.39}$ & $0.048_{-0.032}^{+0.070}$\\
8738735 & $13.9$ & $0.355_{-0.044}^{+0.045}$ & $0.042_{-0.019}^{+0.037}$ & $9.9_{-5.0}^{+14.9}$ & $697.8538_{-0.0049}^{+0.0059}$ & $4.8_{-1.8}^{+4.0}$ & $0.54_{-0.27}^{+0.54}$ & $0.053_{-0.040}^{+0.127}$\\
8800954 & $13.4$ & $0.386_{-0.025}^{+0.025}$ & $0.049_{-0.022}^{+0.039}$ & $1.9279957_{-0.0000091}^{+0.0000092}$ & $492.7652_{-0.0024}^{+0.0024}$ & $1.420_{-0.028}^{+0.026}$ & $1.39_{-0.62}^{+1.11}$ & $0.72_{-0.32}^{+0.58}$\\
9306307 & $14.0$ & $1.22_{-0.36}^{+0.49}$ & $4.6_{-4.3}^{+97.9}$ & $4.3_{-1.1}^{+3.3}$ & $1191.35648_{-0.00018}^{+0.00018}$ & $2.39_{-0.45}^{+1.09}$ & $106.0_{-99.4}^{+2186.4}$ & $22.4_{-21.1}^{+488.4}$\\
10187159 & $14.4$ & $0.43_{-0.13}^{+0.21}$ & $0.061_{-0.036}^{+0.096}$ & $4.9_{-1.8}^{+7.6}$ & $604.1102_{-0.0031}^{+0.0023}$ & $2.70_{-0.69}^{+2.35}$ & $1.25_{-0.82}^{+2.42}$ & $0.26_{-0.21}^{+0.69}$\\
10287723 & $13.4$ & $0.266_{-0.024}^{+0.027}$ & $0.026_{-0.012}^{+0.022}$ & $4.9_{-1.3}^{+4.3}$ & $393.5976_{-0.0029}^{+0.0031}$ & $2.58_{-0.46}^{+1.35}$ & $0.62_{-0.28}^{+0.52}$ & $0.117_{-0.068}^{+0.129}$\\
10321319 & $11.9$ & $0.163_{-0.037}^{+0.046}$ & $0.0120_{-0.0058}^{+0.0115}$ & $5.5_{-2.1}^{+8.1}$ & $554.3562_{-0.0063}^{+0.0064}$ & $2.93_{-0.81}^{+2.44}$ & $0.23_{-0.12}^{+0.25}$ & $0.041_{-0.031}^{+0.069}$\\
10602068 & $14.9$ & $2.00_{-0.35}^{+0.66}$ & $162.4_{-161.6}^{+73.9}$ & $3.16_{-0.83}^{+2.65}$ & $830.80892_{-0.00015}^{+0.00015}$ & $2.11_{-0.40}^{+1.09}$ & $3100.4_{-3083.7}^{+1797.2}$ & $889.4_{-885.4}^{+890.8}$\\
10842718 & $14.6$ & $0.74_{-0.16}^{+0.16}$ & $0.19_{-0.12}^{+22.15}$ & $12.7_{-6.6}^{+20.2}$ & $226.2344_{-0.0047}^{+0.0047}$ & $5.3_{-2.1}^{+4.7}$ & $2.6_{-1.7}^{+324.0}$ & $0.25_{-0.22}^{+30.41}$\\
11709124 & $14.5$ & $0.83_{-0.11}^{+0.12}$ & $0.93_{-0.82}^{+35.44}$ & $4.3_{-1.3}^{+4.7}$ & $657.2674_{-0.0016}^{+0.0018}$ & $2.54_{-0.56}^{+1.62}$ & $18.1_{-16.0}^{+668.5}$ & $3.5_{-3.1}^{+146.2}$\\
\enddata

\tablecomments{The masses and radial velocity amplitudes are estimated based
on the measured radius using a probabilistic mass--radius relation
\citep{Chen:2016}.}
\end{deluxetable}
\end{floattable}

\begin{figure*}[p]~\\
\begin{center}
\includegraphics[width=\textwidth]{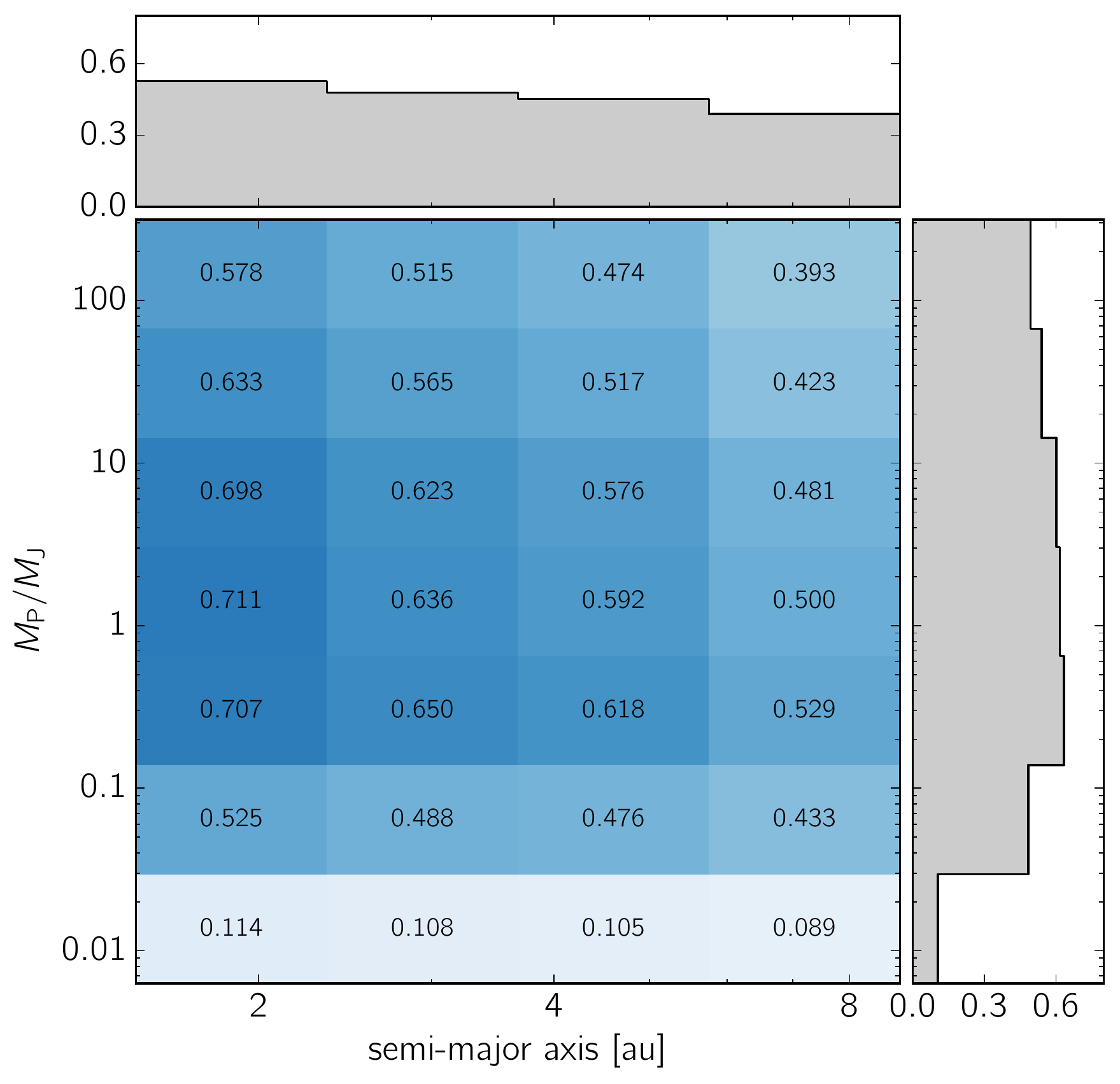}
\end{center}
\caption{%
The same as \dfmfig{completeness} converted into the planet mass and
semi-major axis plane.
Since the completeness function depends on the planet's radius and not its
mass, a probabilistic mass--radius relationship \citep{Chen:2016} was used to
convert radius to mass.
\dfmfiglabel{completeness-am}}
\end{figure*}

\begin{floattable}
\begin{deluxetable}{ccc}
\tabletypesize{\footnotesize}
\caption{The occurrence rate computed in mass units
\label{tab:occrates}}

\tablehead{%
    \colhead{volume} & \colhead{rate density\tablenotemark{a}} & \colhead{integrated rate}
}

\startdata
$2\unit{yr}<P<25\unit{yr};\,0.01\,M_\mathrm{J} \le M < 20\,M_\mathrm{J}$ &
\massocc & \massoccint \\
$1.5\unit{au}<a<9\unit{au};\,0.01\,M_\mathrm{J} \le M < 20\,M_\mathrm{J}$ &
\masssemiocc & \masssemioccint \\
\enddata

\tablenotetext{a}{The rate density is measured in natural logarithmic units;
see \eq{massocc}.}
\tablecomments{These values are computed assuming that the occurrence rate is
flat in the logarithmic parameters.}
\end{deluxetable}
\end{floattable}

\section{Prospects for follow-up}\sectlabel{follow-up}

A real concern about the detection of exoplanets from a single transit is that
follow-up and confirmation is difficult.
Since the period of the orbit is poorly constrained and transits are sparse,
any prediction of a subsequent transit time will be too uncertain to schedule
targeted photometric follow-up \citep{Beichman:2016, Dalba:2016}.
Instead, follow-up using radial velocity and astrometry are more promising.
For both radial velocity and astrometry, there is information about the
orbiting planets in measurements made at all times~--~not just during transit.
This allows observations to be scheduled without a well constrained orbital
period.
Furthermore, follow-up of any of these candidates using radial velocity or
astrometry would provide a measurement of the density of a planet that would
be valuable for the study of planetary compositions.

Table~\ref{tab:masses} lists the posterior predictions for the semi-amplitude
$K$ of the radial velocity signal produced by each candidate using the mass
predictions from the previous section.
Since the orbital periods are long, we also include a simple prediction for
the slope of the radial velocity trend induced by this planet by taking the
ratio of the semi-amplitude and the orbital period.
Any radial velocity follow-up of the candidates presented here would be an
ambitious undertaking because the stars are relatively faint and, in most
cases, the radial velocity trends are small.
Some candidates should, however, be within the reach of current
state-of-the-art facilities.

In principle the \gaia\ Mission will be very sensitive to the astrometric
wobble produced by a long-period exoplanet \citep{Perryman:2014}.
To leading order, the astrometric signal strength is proportional to the
semi-major axis of the stellar (primary) reflex motion in angular units.
That is, detectability is related to the angle $\alpha$ given by
\begin{eqnarray}
\alpha &=& \frac{a}{D}\,\frac{M_{\mathrm p}}{M_{\mathrm s}} \quad ,
\end{eqnarray}
where $a$ is the semi-major axis, $D$ is the distance from the observer to the
primary, and $M_{\mathrm{p}}/M_{\mathrm{s}}$ is the planet-to-star mass ratio.

The single-visit precision of \gaia\ will vary with magnitude but is expected
to be on the order of 40\,$\mu$as at these magnitudes.
In detail the confidence with which an exoplanet can be detected or measured
in the final \gaia\ data depends on this precision, the number of crossings of
the star through the \gaia\ field-of-view, and details of how the projected
orbit is sampled by the time history of the focal-plane crossings.
However, it is not expected that \gaia\ can detect or precisely measure
exoplanet-induced astrometric wobbles that are much smaller in amplitude than
the single-visit precision \citep{Perryman:2014}.

The primary stars in the \kepler\  Field are typically at distances of $\sim
500$\,pc, and typical mass ratios are in the $10^{-4}$ range.
We therefore expect astrometric amplitudes in the 0.3 to 3\,$\mu$as range.
These planets will not be detectable or measurable in the \gaia\ data under
any circumstances, but it may be possible to identify which candidates are
false-alarms caused by eclipsing binaries.
However, similar planets around closer stars \emph{will} be detectable with
\gaia.
This means that there will be a comparable exoplanet occurrence rate
measurement from the \gaia\ data.
It also means that many of the discoveries of the \KT\ and \tess\ missions
could be followed up and precisely measured by the \gaia\ Mission.

\section{Summary}\sectlabel{summary}

We have developed a fully automated method to search for the transits of
long-period exoplanets with only one or two observable transits in the
\kepler\ archival light curves.
This method uses probabilistic model comparison to veto non-transit signals.
Applying this method to the brightest \numtargets\ G/K dwarfs in the \kepler\
target list, we discover \numcands\ systems with likely astrophysical
transits and eclipses.
We fit the light curve for each candidate with a physical generative model and
informative priors on eccentricity and stellar density to estimate the
planet's orbital period.
The constraint on the period is also informed by the simplifying assumption
that no other transit could occur during the baseline of \kepler\ observations
of the target.
Simulations of the false positive population~--~lone primary or secondary
eclipses of binary systems or background eclipsing binaries~--~suggest that 13
of these candidates have high probability of being planetary in nature.

We measure the empirical detection efficiency function of our search procedure
by injecting simulated transit signals into the target light curves and
measuring the recovery rate.
By combining the measured detection efficiency and the catalog of exoplanet
candidates, we estimate the integrated occurrence rate of exoplanets with
orbital periods in the range $2-25\unit{years}$ and radii in the range
$0.1-1\,R_\mathrm{J}$ to be \intocc\ planets per G/K dwarf.
\response{%
This result is qualitatively consistent with estimates of the occurrence rate
of long-period giant planets based on data from radial velocity and direct
imaging surveys.
The occurrence rate measured here~--~for Sun-like hosts~--~is higher than
microlensing results for generally lower mass stars \citep{Gaudi:2012,
Clanton:2014, Clanton:2016} but this discrepancy is consistent with
predictions from the core-accretion model \citep{Laughlin:2004}.
}

Using a probabilistic mass--radius relationship, we predict the masses of our
candidates and report predictions for the radial velocity semi-amplitudes.
Unfortunately, since the target stars are faint and the amplitudes are small,
these targets are unlikely to be accessible with even the current
state-of-the-art high-precision instruments.
We also discuss the potential for astrometric follow-up using the forthcoming
data from the \gaia\ Mission with similarly discouraging results.

Any detailed analysis of individual systems detected with only a single
transit requires follow-up observations to convincingly rule out false
positive scenarios and to better characterize the stellar host parameters
(with, for example, parallax measurements from \gaia).
The conclusions of this work~--~and all other occurrence rate results based on
\kepler\ data~--~are conditioned on the assumption that the stellar
characterization of the target sample is systematic and un-biased.
The main population-level results should be fairly insensitive systematic
issues with the sample but a rigorous analysis of these effects will be
required to come to more detailed conclusions about this population of
long-period transiting planets.

Our method of transit discovery is especially relevant for future photometric
surveys like \KT, \tess, and \plato\ where the survey baseline is shorter than
\kepler.
The transits of planets with orbital periods longer than the observation
baselines will be plentiful in these forthcoming data sets and this method
can, in principle, be trivially generalized to discover these planets,
prioritize follow-up, and study their population.

\vspace{1.5em}
All of the code used in this project is available from
\url{https://github.com/dfm/peerless} under the MIT open-source software
license.
This code (plus some dependencies) can be run to re-generate all of the
figures and results in this \paper; this version of the paper was generated
with git commit \texttt{\githash} (\gitdate).
The parameter estimation results represented as MCMC samplings and the
injection results are available for download from Zenodo at \datareleaseurl.

\vspace{1.5em}
It is a pleasure to thank
Jeff Coughlin,
So Hattori,
Heather Knutson,
Phil Muirhead,
Darin Ragozzine,
Hans-Walter Rix,
Dun Wang,
and
Angie Wolfgang
for helpful discussions and contributions.
We thank the anonymous referee for comments that improved the presentation and
clarity of this manuscript.

T.D.M.\ was supported by the National Aeronautics and Space
Administration, under the \kepler\ participating
scientist program (grant NNX14AE11G), and is grateful for the
hospitality of both the Institute for Advanced Study and Carnegie
Observatories that helped support this work.
D.W.H.\ was partially supported by the National Science Foundation (grant
IIS-1124794), the National Aeronautics and Space Administration (grant
NNX12AI50G), and the Moore--Sloan Data Science Environment at NYU.
E.A.\ acknowledges support from NASA grants NNX13AF20G, NNX13AF62G, and NASA
Astrobiology Institutes Virtual Planetary Laboratory, supported by NASA under
cooperative agreement NNH05ZDA001C.

This research made use of the NASA \project{Astrophysics Data System} and the
NASA Exoplanet Archive.
The Exoplanet Archive is operated by the California Institute of Technology,
under contract with NASA under the Exoplanet Exploration Program.

This \paper\ includes data collected by the \kepler\ mission. Funding for the
\kepler\ mission is provided by the NASA Science Mission directorate.
We are grateful to the entire \kepler\ team, past and present.
Their tireless efforts were all essential to the tremendous success of the
mission and the successes of \KT, present and future.

These data were obtained from the Mikulski Archive for Space Telescopes
(MAST).
STScI is operated by the Association of Universities for Research in
Astronomy, Inc., under NASA contract NAS5-26555.
Support for MAST is provided by the NASA Office of Space Science via grant
NNX13AC07G and by other grants and contracts.

Computing resources were provided by High Performance Computing at New York
University.

\facility{Kepler}
\software{%
    \project{batman} \citep{Kreidberg:2015},
    \project{ceres} \citep{Agarwal:2016},
    \project{corner.py} \citep{Foreman-Mackey:2016},
    \project{emcee} \citep{Foreman-Mackey:2013},
    \project{exosyspop} \citep{Morton:2016a},
    \project{george} \citep{Ambikasaran:2016},
    \project{isochrones} \citep{Morton:2015},
	\project{matplotlib} \citep{Hunter:2007},
	\project{numpy} \citep{Van-Der-Walt:2011},
	\project{scipy} \citep{Jones:2001},
    \project{transit} \citep{Foreman-Mackey:2016a},
  \project{vespa} \citep{Morton:2015b}}.

\newpage
\appendix

\section{Details of the light curve models}\sectlabel{model-details}

In \sect{light-curve-vetting}, the five light curve models were listed.
In this section, we give the mathematical details of each model and list the
parameters that are fit.
Each model~--~except the \modelname{transit} model~--~can be easily
differentiated with respect to its parameters.
As discussed in the following section, this feature is crucial for efficient
and robust likelihood maximization.

\begin{itemize}

{\item
The \modelname{box} model is given by
\begin{eqnarray}
m_\mathrm{box}(t) &=& \left\{\begin{array}{ll}
a, & \mbox{if $t \le t_\mathrm{min}$} \\
b, & \mbox{if $t_\mathrm{min} < t \le t_\mathrm{max}$} \\
c, & \mbox{if $t_\mathrm{max} < t$}
\end{array}\right.
\end{eqnarray}
where $a$, $b$, and $c$ are free parameters, and $t_\mathrm{min}$ and
$t_\mathrm{max}$ are fixed.
In practice, we include two different \modelname{box} models where
$t_\mathrm{min}$ and $t_\mathrm{max}$ are set using different heuristics.
The first \modelname{box} has the bounds set to match the ingress and
egress of the best fit \modelname{transit}.
The second \modelname{box} is chosen based on the largest change points in the
light curve.
}

{\item
The \modelname{step} model is given by
\begin{eqnarray}
m_\mathrm{step}(t) &=& \left\{\begin{array}{ll}
m_1 + h_1\,\exp\left([t - t_0] / w_1\right), & \mbox{if $t < t_0$} \\
m_2 + h_2\,\exp\left([t_0 - t] / w_2\right), & \mbox{if $t_0 \le t$}
\end{array}\right.
\end{eqnarray}
where all of the parameters~--~including $t_0$~--~are included in the fit.
To ensure that the widths $w_1$ and $w_2$ remain positive, we fit for
$\log w_1$ and $\log w_2$.
}

{\item
For the \modelname{outlier} model, we iterate through all cadences $t_n$
within 0.3~days of the candidate transit time and evaluate the model as
\begin{eqnarray}
m_\mathrm{outlier}(t) &=& \left\{\begin{array}{ll}
f(t_0), & \mbox{if $t = t_0$} \\
\mathrm{median}[f(t \ne t_0)], & \mbox{if $t \ne t_0$}
\end{array}\right.
\end{eqnarray}
where $f(t)$ is the observed time series.
With this model, no non-linear optimization is required and the final value of
$t_0$ is the one with the maximum likelihood in this grid search.
}

{\item
The \modelname{variability} model only has one parameter, the flux $m_0$ and
$m_\mathrm{variability}(t) = m_0$ at all times.
The variability is captured by the Gaussian Process residual model.
}

{\item
Finally, the \modelname{transit} model is an exposure time integrated,
limb darkened light curve \citep{Mandel:2002, Kipping:2010} parameterized by
the radius ratio between the planet and star, the transit duration, the
transit time, the impact parameter, and two quadratic limb darkening
coefficients \citep{Kipping:2013}.
Analytically computing the gradient of a simple transit model is possible
\citep{Pal:2008} but it becomes substantially more tedious as the model
becomes more realistic.
Therefore, we instead use a compile-time automatic differentiation
library\footnote{More specifically, we use the \texttt{Jet} object from the
BSD-licensed Ceres Solver \url{http://ceres-solver.org}} \citep{Agarwal:2016}
to efficiently compute first derivatives of the full transit model with
respect to the orbital and physical parameters to machine precision.
}

\end{itemize}

\section{Gaussian process regression}\sectlabel{gp-regression}

Gaussian Processes (GPs) are a class of non-parametric, stochastic models that
have been demonstrated to be good effective models for the variability in
\kepler\ light curves.
A simple GP model can be used to capture residual non-transit variability in
light curves.
In this \paper, we use a GP model for two steps: light curve--level transit
shape vetting and parameter estimation.
A full discussion of GPs is beyond the scope of this \paper, so we will only
summarize the most relevant points here and direct an interested reader to
\citet{Rasmussen:2006} for more details.

A GP model is specified by the following likelihood function
\begin{eqnarray}\eqlabel{gplike}
\mathcal{L} = \ln p(\bvec{y}\,|\,\meanpars,\,\kernpars) &=&
- \frac{1}{2}\,\bvec{r}(\meanpars)^\T\,K(\kernpars)^{-1}\,
    \bvec{r}(\meanpars)
- \frac{1}{2}\log\det K(\kernpars) - \frac{N}{2} \log{2\,\pi}
\end{eqnarray}
where \bvec{y} is a list of measurements in a scalar time series~--~in this
case, fluxes~--~measured at the times \bvec{t}, and
\begin{eqnarray}
\bvec{r}(\meanpars) &=& \bvec{y} - m(\bvec{t};\,\meanpars)
\end{eqnarray}
is the vector of residuals away from the mean model $m(\bvec{t};\,\meanpars)$.
For the purposes of this paper, we model the covariance matrix $K(\kernpars)$
using the Mat\'ern-3/2 kernel.
Under this model, the elements of $K(\kernpars)$ are given by
\begin{eqnarray}\eqlabel{matern}
\left[ K(\kernpars) \right]_{ij} &=& \sigma_i^2\,\delta_{ij}
    + \alpha^2 \left[ 1+\frac{|t_i - t_j|}{\sqrt{3}\,\tau} \right]
      \exp \left(-\frac{|t_i - t_j|}{\sqrt{3}\,\tau}\right)
\end{eqnarray}
where $\sigma_i$ is the reported uncertainty on the $i$-th measurement in the
time series and $\delta_{ij}$ is the Kronecker delta.

This covariance function (\eqalt{matern}) is specified by an amplitude
$\alpha$ and a time scale $\tau$ and we will simultaneously fit for these
hyperparameters $\kernpars=(\alpha,\,\tau)$ and the parameters of the mean
model \meanpars.
To efficiently find the parameter set that maximizes \eq{gplike} using a
non-linear optimization routine\footnote{We use the L-BFGS-B method as
implemented in SciPy \url{%
http://docs.scipy.org/doc/scipy/reference/generated/%
scipy.optimize.minimize.html}.},
it is useful to be able to compute the gradient of \eq{gplike} with respect to
the parameters \meanpars\ and \kernpars.
These gradients are given by
\begin{eqnarray}\eqlabel{gpmeangrad}
\frac{\dd\ln p(\bvec{y}\,|\,\meanpars,\,\kernpars)}{\dd \meanpars} &=&
\frac{\dd m(\bvec{t};\,\meanpars)}{\dd\meanpars}^\T \, K(\kernpars)^{-1} \,
    \bvec{r}(\meanpars)
\end{eqnarray}
and
\begin{eqnarray}
\frac{\dd\ln p(\bvec{y}\,|\,\meanpars,\,\kernpars)}{\dd \kernpars} &=&
\frac{1}{2}\,\mathrm{Tr}\left(
    \left[ \bvec{\phi}\,\bvec{\phi}^\T - K(\kernpars)^{-1} \right]
    \,\frac{\dd K(\kernpars)}{\dd\kernpars}
\right)
\end{eqnarray}
where
\begin{eqnarray}
\bvec{\phi} &=& K(\kernpars)^{-1}\,\bvec{r}(\meanpars) \quad.
\end{eqnarray}

\bibliography{peerless}

\end{document}